\documentclass[letterpaper,12pt]{article}
\usepackage[margin=1in,letterpaper]{geometry} % decreases margins

\usepackage{graphicx}% Include figure files
\usepackage{dcolumn}% Align table columns on decimal point
\usepackage{bm}% bold math
\usepackage{hyperref}% add hypertext capabilities
\usepackage{soul}
\usepackage{xcolor}
\usepackage{amsmath}
\usepackage{amssymb}
\usepackage{tabularx}
\usepackage{makecell}
\usepackage[normalem]{ulem}

\usepackage{multirow}

\newcommand{\rvec}{\vec{\mathbf{r}}}
\newcommand{\rvecp}{\vec{\mathbf{r}}\,'}
\newcommand{\Gammath}{\langle\Gamma_\text{th}\rangle}
\newcommand{\cE}{\mathcal{E}}

\begin{document}

\section*{ }
\begin{minipage}{\textwidth}
    \LARGE
    \centering
    Single-photon superradiance and subradiance in helical collectives of quantum emitters
    \vspace{10mm}
\end{minipage}

\begin{minipage}{\textwidth}
    \large
    \centering
    Hamza Patwa\,$^{1}$ and Philip Kurian\,$^{1*}$
    \vspace{10mm}
\end{minipage}

\begin{minipage}{\textwidth}
    \centering
    $^{1}$Quantum Biology Laboratory, Howard University, Washington, D.C. 20060, USA\\
    \url{https://www.quantumbiolab.com}
    \vspace{10mm}
\end{minipage}

\begin{minipage}{\textwidth}
    \centering
    *E-mail: pkurian@howard.edu
    \vspace{5mm}
\end{minipage}

\begin{abstract}
    Collective emission of light from distributions of two-level systems was first predicted in 1954 by Robert Dicke, who showed that when $N$ quantum emitters absorb photons, their collective radiative decay rate can be significantly enhanced (superradiance) or suppressed (subradiance) relative to the single-emitter decay rate. In this work, we derive novel analytical expressions for the collective decay rates and collective Lamb shifts for the coherent interaction of a single photon with topologically one-dimensional, continuous distributions of quantum two-level systems: an infinite line and an infinite helix. We compare these solutions to higher-dimensional collectives of two-level systems (cylinder), finding certain limits in which the eigenvalues of structures of different dimensions become equal. We also compare our solution with arrangements where the distribution of transition dipole vector emitters is discrete rather than continuous, and when short- ($1/r^3$), intermediate- ($1/r^2$), and long-range ($1/r$) interaction terms are included. We find important differences between the discrete vector and continuous scalar emitter cases, which do not agree in the limit where discrete spacing goes to 0. The analytical solution for the helix is then used to make order-of-magnitude estimates of the maximally superradiant state, thermally averaged collective decay rate, and percentage of trapped states in helical architectures of molecular quantum emitters in protein fibers. Given the differences in inclusion of short- and intermediate-range interaction terms between the numerical and analytical models for realistic protein fiber architectures, these results show excellent agreement for sparse arrangements of emitters. Our work thus bridges the theoretical gap between different formalisms for treating superradiant matter distributions, aids the engineering of helical devices which harness quantum optical effects for computing with superradiant error correction and subradiant memories, and motivates the discovery and creation of flexible platforms for quantum information processing using the intrinsic helical geometries of biomatter.
\end{abstract}

\section{\label{sect:intro}Introduction}
The phenomenon of collective spontaneous emission of light was first discovered by Robert Dicke in 1954 \cite{Dicke1954}. It can lead to an enhancement of the decay rate $\Gamma$ such that it is much larger than the single system decay rate $\gamma$, which is called superradiance, or a supression of the decay rate $\Gamma$ to be much less than $\gamma$, which is called subradiance. This enhancement/supression of the decay rate arises from the interaction of a collective of quantum systems with the environment. Thus, the theoretical formalism that describes superradiance is given frequently in the language of open quantum systems. In this work, we consider the single-photon limit, in which only a single excitation is shared coherently across the collective.

Superradiance (and subradiance) has been studied in many different architectures, and with many different mathematical approaches. Discrete arrangements of two-level systems, such as a line or a lattice, were studied in Ref.~\cite{AAG2017}, where each two-level system is treated as a transition dipole vector, and the interaction of the network of dipoles with the electromagnetic field is modeled with a Lindblad equation. In more realistic structures, such as architectures of quantum emitters in biology, superradiance has also been explored \cite{Patwa2024,Babcock2024} using non-Hermitian quantum mechanics, a formalism for open quantum systems. This formalism is very similar to the one used in Ref.~\cite{AAG2017}. 

In this work, we utilize the approach from Ref.~\cite{Svidzinsky2016}, which deals with \textit{continuous} distributions of matter interacting with a single photon. In going to the continuous limit, closed-form eigensolutions of more complicated idealized structures can be obtained. Closed-form analytical solutions can also be used to guide and inform computational and experimental studies of superradiance and subradiance, which are very time consuming and/or expensive for large structures. We derive in this work novel eigensolutions in the continuous limit for two structures: the infinite continuous line of two-level systems, and an infinite continuous single-helix of two-level systems. The outline of the rest of the paper is the following. In Section \ref{sect:methods}, we outline the formalism from Ref.~\cite{Svidzinsky2016} and some previous results obtained with this formalism. In Sections \ref{sect:infinite_line} and \ref{sect:infinite_helix}, we present our solutions for the infinite line and helix, respectively. In Section \ref{sect:comparison}, we compare our solutions to the ones obtained from different formalisms, and outline why differences appear between them. In Section \ref{tab:estimates_biostructures}, we present estimates of the superradiance and subradiance in realistic protein fibers using the helix eigenvalues. Finally, we conclude and briefly discuss future work in Section \ref{sect:conclusion}.

\section{\label{sect:methods}Methods}
Consider a cloud or collective of identical two-level systems contained in a volume $V$ with a density $n(\rvec)$, each one with an excitation energy $\Delta E=\hbar\omega_0=\hbar c k_0 $, where $k_0\equiv\omega_0/c$ is the wavenumber and $c$ is the speed of light. Every individual system has a decay rate $\gamma$. The system is characterized by a wavefunction $\beta(t,\rvec)$, the norm of which gives the probability amplitude of finding the excitation at the position $\rvec$ at time $t$. The time-evolution of this probability amplitude is given by the following \cite{Svidzinsky2016}:
\begin{equation}
    \label{eq:time-evolution}
    \frac{\partial \beta(t,\rvec)}{\partial t} = i\gamma\int_V d\rvecp n(\rvecp) \beta(t,\rvecp) \frac{\exp (ik_0 |\rvec - \rvecp|)}{k_0 |\rvec - \rvecp|}
\end{equation}
The solutions to this equation are of the following form
\begin{equation}
    \label{eq:form-of-beta}
    \beta(t,\rvec)=e^{-\mathcal{E} t}\beta(\rvec),
\end{equation}
where $\mathcal{E} \in \mathbb{C}$ and $\beta(\rvec)$ satisfies an eigenvalue equation
\begin{equation}
    \label{eq:eigenvalue-eq}
    \mathcal{E} \beta(\rvec) = -i\gamma\int_V d\rvecp n(\rvecp) \beta(\rvecp) \frac{\exp (ik_0 |\rvec - \rvecp|)}{k_0 |\rvec - \rvecp|}.
\end{equation}
The eigenvalue $\mathcal{E}$ can be decomposed into real and imaginary parts
\begin{equation}
    \mathcal{E}=\frac{\Gamma}{2}+iE
    \label{eq:real_and_imag_parts_of_eigenvalue_general}
\end{equation}
where the real part $\Gamma$ is the collective decay rate, and the imaginary part $E$ is the collective Lamb shift of the photon from the single-system excitation energy. Note that in other works \cite{Patwa2024,Babcock2024,Celardo2019,AAG2017}, the opposite convention $\cE=E-i\Gamma/2$ is used, replacing the exponential argument in Eq.~\eqref{eq:form-of-beta} with $-i\cE$. When we compare results from different formalisms, this is taken into account. In Ref.~\cite{Svidzinsky2016}, the above equations \eqref{eq:time-evolution}-\eqref{eq:eigenvalue-eq} have been solved for an infinite cylinder, a sphere, and a spheroid with two-level systems distributed on their surfaces. In this work, we add to those results by studying two more systems: an infinite helix and an infinite line with a continuous distribution of two-level systems. Helical geometries are ubiquitous in practical scenarios, such as in biological structures (e.g., proteins and nucleic acids). In fact, superradiance has been studied theoretically \cite{Patwa2024} and confirmed experimentally \cite{Babcock2024} in helical protein fiber architectures.

The eigenvalues for an infinite cylinder were calculated in Ref.~\cite{Svidzinsky2016}: 
\begin{align}
    \Gamma & = \frac{\pi \gamma n_0}{2k_0}J_n^2\left(\sqrt{k_0^2-k_z^2}R\right) \label{eq:svidzinsky_cylinder_Gammas}\\
    E & = \frac{\pi \gamma n_0}{k_0}J_n\left(\sqrt{k_0^2-k_z^2}R\right) Y_n\left(\sqrt{k_0^2-k_z^2}R\right) \label{eq:svidzinsky_cylinder_Energies}
\end{align}
The eigenfunctions are
\begin{equation}
    \label{eq:svidzinsky_cylinder_eigfuncts}
    \beta(\rvec)=\beta(\varphi, z)=e^{in\varphi}e^{ik_zz}.\nonumber
\end{equation}
where \(n \in \mathbb{Z}\) and \(k_z \in \mathbb{R}\) are the quantum numbers in the azimuthal and z-directions, respectively.

The maximally superradiant state for the infinite helix occurs when $k_z=k_0$, while trapped states develop when the argument of the Bessel function of the first kind is a zero of the Bessel function. Trapped states also occur when $k_z>k_0$. The maximally superradiant state also has a collective Lamb shift which diverges to $-\infty$. This is a feature that will show up in all the other systems we study in this work. 

\section{\label{sect:infinite_line}Analytical Solution for Infinite Line of Quantum Emitters}
\begin{figure}
    \centering
    \includegraphics[width=0.6\linewidth]{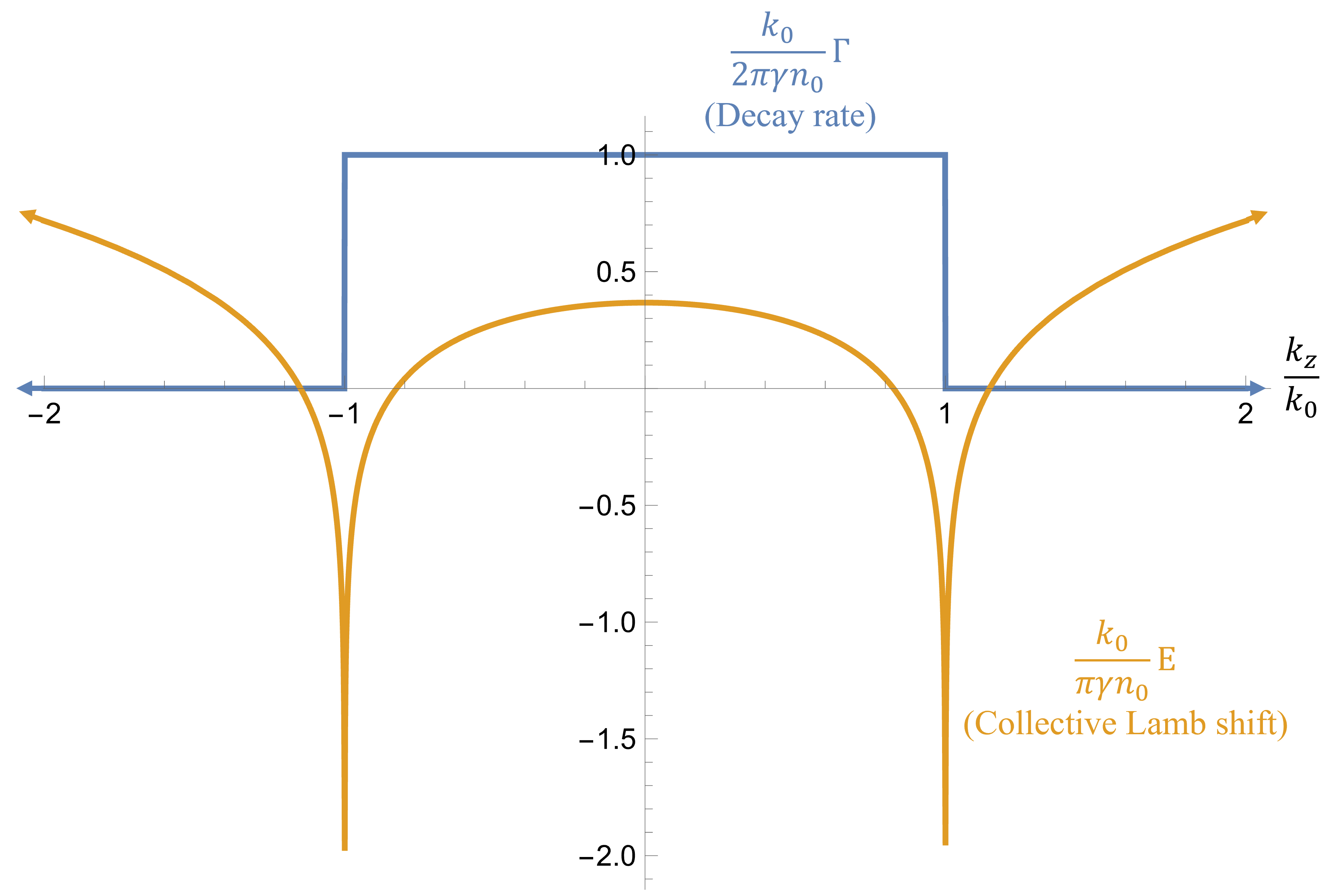}
    \caption{\textbf{An infinite continuous line of quantum emitters exhibits either maximum decay-rate states, or trapped states with exactly zero decay rate.} The blue curve is the plot of Eq.~\eqref{eq:line_gamma} as a function of $\kappa=k_z/k_0$ and the yellow curve is the plot of Eq.~\eqref{eq:line_E} as a function of $\kappa$. Trapped states occur for any $|\kappa|\geq1$, and the other states for $|\kappa|<1$ are all of equal maximal decay rate. The collective Lamb shift of the states at $\kappa=\pm1$ diverges to $-\infty$.}
    \label{fig:infinite_line_vs_kz_k0}
\end{figure}

We consider in this section quantum emitters continuously distributed on an infinitely long line. This system has been studied in the discrete case \cite{AAG2017}, where emitters are equally spaced on an infinite line with finite spacing $d=d_0$. Our system is similar to the system of Ref.~\cite{AAG2017}, but in the limit $d\rightarrow 0$, such that the line density $n_0$ is the same as it was before, i.e., $n_0=1/d_0$. We will compare our solution to their solution in this limit. Let the line be oriented such that it lies along the $z$-axis. The distance $|\rvec-\rvec'|$ in the eigenvalue equation \eqref{eq:eigenvalue-eq} then simplifies to a scalar expression $|z-z'|$, giving us
\begin{equation}
    \mathcal{E}\,\beta(z)=\frac{-i \gamma n_0}{k_0}\int_{-\infty}^{\infty} dz'\,\frac{\exp{\left( ik_0\left|z-z'\right| \right)}}{\left|z-z'\right|} \,\beta(z')\nonumber
\end{equation}
Changing variables inside the integral to $u\equiv k_0(z'-z)$, and defining $K(u)$ as the kernel,
\begin{equation}
    \mathcal{E}\,\beta(z)=\frac{-i \gamma n_0}{k_0}\int_{-\infty}^{\infty} du\,K(u) \,\beta(u/k_0+z)\nonumber
\end{equation}
with
\begin{equation}
    K(u)\equiv \frac{\exp{\left( i\left|u\right| \right)}}{\left|u\right|}.\nonumber
\end{equation}
We can then replace the kernel $K$ with its Fourier representation. The Fourier transform of $K$ is $\widetilde{K}$, where
\begin{align}
    \widetilde{K}(\omega)=-2\gamma_E-\ln\left|\omega^2-1\right|+i\pi\Theta(1-|\omega|)\label{eq:FT_of_K}
\end{align}
where $\gamma_E$ is the Euler gamma constant. $\Theta(x)$ is the step function, defined to be $1$ when its argument is positive, and $0$ when its argument is $0$ or negative. The argument $\omega$ is dimensionless. The eigenvalue equation then reads
\begin{align}
    \mathcal{E}\,\beta(z)=\frac{-i \gamma n_0}{2\pi k_0}\int_{-\infty}^{\infty} du\, \int_{-\infty}^{\infty}d\omega\,\widetilde{K}(\omega)\,e^{i\omega u} \,\beta(u/k_0+z)\nonumber
\end{align}
Now, we will use the ansatz $\beta(z)=e^{i k_z z}$, where $k_z$ is the quantum number parameterizing the eigenfunction. It will also appear in the eigenvalue expression. Plugging this in,
\begin{align}
    \mathcal{E}\,e^{i k_z z}=\frac{-i \gamma n_0}{2\pi k_0}\int_{-\infty}^{\infty} du\, \int_{-\infty}^{\infty}d\omega\,\widetilde{K}(\omega)\,e^{i\omega u} \,e^{i k_z (u/k_0+z)}\nonumber
\end{align}
We now switch the order of integration and take all factors not dependent on $u$ or $\omega$ out of the integrals, obtaining
\begin{align}
    \mathcal{E}\,e^{i k_z z}=\frac{-i \gamma n_0}{2\pi k_0}e^{i k_z z}\int_{-\infty}^{\infty} d\omega\,\widetilde{K}(\omega) \int_{-\infty}^{\infty}du\,\,e^{i(k_z/k_0+\omega)u} \nonumber
\end{align}
We can now see that the ansatz $\beta(z)=e^{ik_zz}$ was correct, since $e^{ik_zz}$ on the RHS was factored out. The integral over $u$ is a delta function, which collapses the integral over $\omega$ and replaces $\omega$ with $k_z/k_0$. The delta function also comes with a factor of $2\pi$, so the resulting expression is
\begin{align}
    \mathcal{E}\,e^{i k_z z}=\frac{-i \gamma n_0}{k_0}\widetilde{K}(k_z/k_0)\,e^{i k_z z}\nonumber
\end{align}
from which we can read off the eigenvalues as
\begin{equation}
    \label{eq:line_eigvals}
    \mathcal{E} = \frac{-i \gamma n_0}{ k_0}\widetilde{K}(k_z/k_0)
\end{equation}
where $\widetilde{K}$ is given by Eq.~\eqref{eq:FT_of_K}. The decay rates $\Gamma$ and collective Lamb shifts $E$ are given by applying Eq.~\eqref{eq:real_and_imag_parts_of_eigenvalue_general} to Eq.~\eqref{eq:line_eigvals}:
\begin{align}
    \frac{k_0}{2\pi\gamma n_0}\Gamma&= \begin{cases}
        1 &|\kappa| \leq 1\\
        0 &|\kappa|>1
    \end{cases}\label{eq:line_gamma}\\
    \frac{k_0}{\gamma n_0}E&=-2\gamma_E-\ln\left|1-\kappa^2\right|
    \label{eq:line_E}
\end{align}
where $\kappa=k_z/k_0$. We can see that the decay rates are maximal and the same for $|\kappa|<1$, and trapped for $|\kappa|\geq1$. The collective Lamb shifts diverge for $\kappa=\pm1$. \textcolor{black}{Eqs.~\eqref{eq:line_gamma} and \eqref{eq:line_E} are plotted in Fig.~\ref{fig:infinite_line_vs_kz_k0}, which shows these features.}
\section{\label{sect:infinite_helix}Analytical solution for Infinite Helix of Quantum Emitters}
We consider in this section a helical structure that can be parametrized in cylindrical coordinates $(\rho,\theta,z)$ as
\begin{equation}
    \label{eq:6.13}
    \begin{cases}
    \rho=R\\
    \theta=\frac{2\pi}{b} z\\
    z=z
    \end{cases}\nonumber
\end{equation}
where \(R\) is the constant radius of the helix. \(z\) is the vertical distance, and it also functions as the parametrization variable. $b$ is the pitch of the helix, which represents the spacing between adjacent rings on the helix. We consider a constant line density \(n(\rvec)=n_0\) on this helix, where \(\rvec\) is a vector that lies on the helix. We can then simply make the substitution \(\theta\rightarrow 2\pi z/b\) into the eigenvalue equation \eqref{eq:eigenvalue-eq} written in cylindrical coordinates, resulting in a one-dimensional problem:
\begin{align}
    \mathcal{E}\,\beta( z)=\frac{-i \gamma n_0}{k_0}\int_{-\infty}^{\infty} dz'\,K(z,z') \,\beta(z')\nonumber
\end{align}
where in this case, the kernel $K$ is a function of $u\equiv z'-z$,
\begin{align}
    K(u)\equiv\frac{\exp\left(ik_0\sqrt{2R^2-2R^2\cos(2\pi u/b)+u^2}\right)}{\sqrt{2R^2-2R^2\cos(2\pi u/b)+u^2}}\nonumber
\end{align}
We use the following expansion for the kernel $K$:
\begin{align}
    K(u)=\frac{i}{2}\int_{-\infty}^{\infty}\sum_{m=-\infty}^{\infty} J_m\left(\sqrt{k_0^2-k^2}R\right)\nonumber H_m^{(1)}\left(\sqrt{k_0^2-k^2}R\right)e^{i(2\pi m/b+k)u}\nonumber
\end{align}
where $J_m(x)$ is the Bessel function of the first kind and $H_m^{(1)}(x)$ is the Hankel function of the first kind. Substituting this expansion, the eigenvalue equation becomes
\begin{align}
    \mathcal{E}\,\beta(z)&=\frac{\gamma n_0}{2k_0} \int_{-\infty}^\infty dz'\int_{-\infty}^\infty dk \sum_{m=-\infty}^{\infty} J_m\left(\sqrt{k_0^2-k^2}R\right)\nonumber\\&\times H_m^{(1)}\left(\sqrt{k_0^2-k^2}R\right) e^{i(2\pi m/b+k)(z-z')}\beta(z')\nonumber
\end{align}
We can now use the ansatz \(\beta(z)=e^{ik_zz}\). Substituting this into the eigenvalue equation, we obtain
\begin{align}
    \mathcal{E}\,e^{ik_zz}&=\frac{\gamma n_0}{2k_0} \int_{-\infty}^\infty dz'\int_{-\infty}^\infty dk \sum_{m=-\infty}^{\infty} J_m\left(\sqrt{k_0^2-k^2}R\right)H_m^{(1)}(\sqrt{k_0^2-k^2}R) e^{i(2\pi m/b+k)(z-z')}e^{ik_zz'}\nonumber\\
    &=\frac{\pi \gamma n_0}{k_0} \int_{-\infty}^\infty dk \sum_{m=-\infty}^{\infty} J_m\left(\sqrt{k_0^2-k^2}R\right)H_m^{(1)}\left(\sqrt{k_0^2-k^2}R\right)e^{i(2\pi m/b+k)z}\delta(-2\pi m/b-k+k_z)\nonumber\\
    &=\bigg[\frac{\pi \gamma n_0}{k_0}\sum_{m=-\infty}^{\infty} J_m\left(\sqrt{k_0^2-(k_z-2\pi m/b)^2}\,R\right)H_m^{(1)}\left(\sqrt{k_0^2-(k_z-2\pi m/b)^2}\,R\right)\bigg]e^{ik_zz}.\nonumber
\end{align}
In the last equation above, we can see that the eigenfunction \(\beta(z)=e^{ik_z z}\) was correct, since all the \(z\)-dependence is contained in that eigenfunction, and it is the same on both sides. We can then identify the term in brackets on the RHS as the eigenvalue \(\mathcal{E}\), and  redefine all variables inside the Bessel/Hankel function arguments so that they are dimensionless:
\begin{equation}
    \frac{k_0}{\pi \gamma n_0}\cE = \sum_{m=-\infty}^{\infty} J_mH^{(1)}_m\left(\sqrt{1-(\kappa-m\Omega)^2}\,r\right)\label{eq:helix_eigenvalue}
\end{equation}

where the shorthand $J_m H^{(1)}_m(x)\equiv J_m(x)H_m^{(1)}(x)$ is used for brevity. The other adimensional variables are $\kappa=k_z/k_0$, $\Omega=\frac{2\pi}{k_0 b}$, and $r=k_0 R$. The collective wavenumber $k_z$ is normalized by the single-system wavenumber $k_0$ to make $\kappa$, the adimensional inverse pitch is $\Omega$, and the adimensional radius is $r$. To separate the eigenvalue into real and imaginary parts, we can use some information about Bessel and Hankel functions of the first kind. Specifically, we know the following:
\begin{align}
    &\lim_{x\to 0} J_0 H_0^{(1)}(x) =1-i\,\infty\label{eq:fact1}\\
    &\lim_{x\to 0} J_m H_m^{(1)}(x) =0-\frac{i}{|m|\pi},\, m \in \mathbb{Z},\, m \neq 0\label{eq:fact2}\\
    &J_m H_m^{(1)}(x) = a + bi,\,\,\, \,\,\,\,\,\,\,\,\,\,\,\,\,\,\,\, \,\,\,a \in [0,1) \text{, } x\in(0,\infty)\label{eq:fact3}\\
    &J_m H_m^{(1)}(ix) = 0 - ci,\,\,\,\,\,\,\,\,\,\,\,\,\,\, \,\,\,\,\,\,c > 0 \text{, }\,\,\,\,\,\,\,\,\,\, x\in(0,\infty)\label{eq:fact4}
\end{align}
Since the argument of the Bessel/Hankel function is always either real and positive or imaginary and positive, Eqs. (\ref{eq:fact1})-(\ref{eq:fact4}) cover every possible value of the Bessel/Hankel function argument in \eqref{eq:helix_eigenvalue}. Observing the real parts of Eqs. (\ref{eq:fact1})-(\ref{eq:fact4}), we can see that the real part of every term of the sum (\ref{eq:helix_eigenvalue}) is non-negative, and therefore the decay rate is always non-negative, which is what we would expect physically. Now, define two critical integers \(m_{\text{min}}\) and \(m_{\text{max}}\). $m_{\text{min}}$ ($m_{\text{max}}$) is the smallest (largest) integer \(m\) that keeps the Bessel/Hankel function argument real. We can write these as
\begin{equation}
    \label{eq:m_min_and_m_max}
    m_{\text{min}} \equiv \bigg\lceil \frac{\kappa-1}{\Omega} \bigg\rceil,\,\,\,\,m_{\text{max}} \equiv \bigg\lfloor \frac{\kappa+1}{\Omega} \bigg\rfloor
\end{equation}
Note that these definitions are only valid if $m_{\text{min}} \leq m_{\text{max}}$, otherwise there is no integer $m$ that makes the Bessel/Hankel function argument a real number.

All of the terms with \(m\notin [m_{\text{min}},m_{\text{max}}]\) are of type \eqref{eq:fact4}, so they have 0 real part. This means that only the terms inside $[m_{\text{min}},m_{\text{max}}]$ contribute to the decay rate. We can then write an exact expression for the decay rate as a finite sum
\begin{equation}
    \label{eq:helix_eigenvalue_real_part}
    \frac{k_0}{2\pi \gamma n_0}\Gamma \,\,= \sum_{m=m_\text{min}}^{m_\text{max}} J_m^2\left(\sqrt{1-(\kappa-m\Omega)^2}\,r\right).
\end{equation}
Since the real part of $H^{(1)}_m(x)$ is $J_m(x)$, substituting into Eq.~\eqref{eq:helix_eigenvalue} we get the square of the Bessel function inside the sum.

\begin{figure*}
    \centering
    \includegraphics[width=\textwidth]{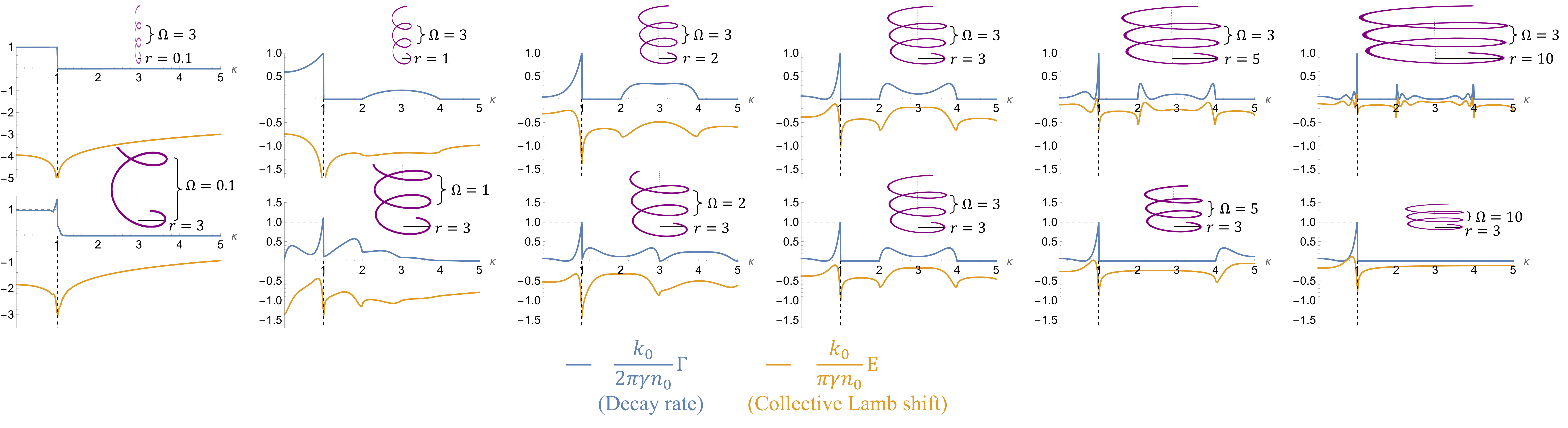}
    \caption{\textbf{Collective Lamb shifts and radiative decay rates for an infinite continuous helix of quantum emitters exhibit clear dependencies on the geometric parameters of inverse pitch ($\Omega = 2\pi/k_0 b$) and radius ($r = k_0R$), with the decay rates approaching those of the infinite continuous line in the limits $r\rightarrow0$ and $\Omega\rightarrow0$.} The blue and orange curves in each plot correspond to Eqs. \eqref{eq:helix_eigenvalue_real_part} and \eqref{eq:helix_eigenvalue_Lamb_shift}, respectively, plotted as a function of $\kappa$ and evaluated at the displayed $\Omega$ and $r$ values. The black vertical dashed lines indicate asymptotes, while the gray horizontal dashed lines are reference grid lines at the vertical coordinate $1.0$. For the collective Lamb shifts, the infinite sum \eqref{eq:helix_eigenvalue_Lamb_shift} is truncated to ten terms for plotting purposes. The top row shows fixed $\Omega=3$ and $r$ varying from $0.1$ to $10$ going left to right. The bottom row shows fixed $r=3$ and $\Omega$ varying from $0.1$ to $10$ going from left to right. In every plot, it is seen that $\kappa=1$ is where the maximum decay rate occurs. The condition \eqref{eq:trapped_states_condition} for trapped states can also be seen in these plots. The decay rates in the two leftmost plots, $\Omega=3,r=0.1$ and $\Omega=0.1,r=3$, approach the decay rates of the infinite continuous line, which can be compared directly in Fig. \ref{fig:infinite_line_vs_kz_k0}.}
    \label{fig:infinite_helix}
\end{figure*}

The collective Lamb shift is an infinite sum:
\begin{equation}
    \label{eq:helix_eigenvalue_Lamb_shift}
    \frac{k_0}{\pi \gamma n_0}E \,\,= \sum_{m=-\infty}^{\infty} J_m Y_m\left(\sqrt{1-(\kappa-m\Omega)^2}\,r\right)
\end{equation}
where the imaginary part of $H_m^{(1)}(x)$, $Y_m(x)$, is substituted and again we use the shorthand $J_mY_m(x)\equiv J_m(x) Y_m(x)$. Although the collective Lamb shift is an infinite sum, we can put an upper bound on it. According to Eq.~\eqref{eq:fact4}, the terms with $m\notin[m_{\text{min}}, m_{\text{max}}]$ all have a strictly negative contribution to the imaginary part of the eigenvalue. Therefore, the sum over $m\in[m_{\text{min}}, m_{\text{max}}]$ will be an upper bound for the collective Lamb shift:
\begin{equation}
    \label{eq:helix_eigenvalue_Lamb_shift_upper_bound}
    \frac{k_0}{\pi \gamma n_0}E \,\,< \sum_{m=m_\text{min}}^{m_\text{max}} J_m Y_m\left(\sqrt{1-(\kappa-m\Omega)^2}r\right)\nonumber
\end{equation}
The decay rates \eqref{eq:helix_eigenvalue_real_part} and collective Lamb shifts \eqref{eq:helix_eigenvalue_Lamb_shift} are plotted in Fig.~\ref{fig:infinite_helix} as a function of $\kappa$ for different helix geometries. For the collective Lamb shifts, since these are given by an infinite sum, we truncate the sum to $21$ terms ($m=-10$ to $m=10$). As the sum is numerically calculated for larger and larger $m$, the difference between this value and the value for 21 terms plateaus to $\sim25\%$ as the terms of the sum drop below machine precision. This difference is very similar for all values of $\kappa$, so increasing the number of terms in the sum would be almost equivalent to adding a constant negative shift to all the collective Lamb shifts, a common tactic employed in non-relativistic quantum mechanical treatments of two-level systems in the electromagnetic field \cite{mattiotti2022,Svidzinsky2016}. Although we currently lack a formal proof that the infinite sum \eqref{eq:helix_eigenvalue_Lamb_shift} converges, the correspondence of the infinite helix decay rates to those of the infinite line in appropriate limits (shown in Section \ref{sect:estimates_biostructures}) serves as an indication that the helix collective Lamb shift values are correct within their (quantum mechanical) domain of applicability. $\Omega$ controls the fraction of states that are strictly trapped (if $\Omega\geq2$, then trapped states with exactly zero decay rate occur in the interval \eqref{eq:trapped_states_condition}), while $r$ controls how much the eigenvalue oscillates with $\kappa$ in the non-trapped regions. In the next section, we elaborate on this statement, outline some properties of the solution, and show how they appear in the plots of Fig.~\ref{fig:infinite_helix}.

\subsection{\label{sect:decay_rate_enhancements}Decay rate enhancements}
Let us now investigate special cases when the decay rate is enhanced. The largest possible term in the sum \eqref{eq:helix_eigenvalue_real_part} will be the $m=0$ term if the argument is also $0$, by Eq. \eqref{eq:fact1}. The value of this term would be $1$. If we set $m=0$ and set the argument $\sqrt{1-(\kappa-m\Omega)^2}\,r=0$, then we immediately obtain $\kappa=\pm1$. In this case, the imaginary part of the eigenvalue diverges to \(-\infty\) by Eq.~\eqref{eq:fact1}, while the real part of this term is 1. So, when $\kappa=\pm1$, $k_0 \Gamma/2\pi \gamma n_0\geq 1$. This represents an enhancement of the decay rate due to resonance between the single-emitter ($k_0$) and collective ($k_z$) wavenumbers. The decay rate enhancement can be seen in all plots in Fig.~\ref{fig:infinite_helix} at $\kappa=1$. So, we have a similarity to the infinite cylinder case in that eigenstates with the maximum decay rate also have a minimum  collective Lamb shift, and that the decay rate is maximized at $\kappa=1$.

\subsection{\label{sec:trapped_states}Trapped states}
Trapped states, for which $\Gamma=0$, would occur if the interval $[m_\text{min},m_\text{max}]$ is null, i.e., $m_\text{min}> m_\text{max}$. In this case, there would be no terms in the sum \eqref{eq:helix_eigenvalue_real_part}, making the decay rate 0. This would happen if $(\kappa, \Omega)$ take values such that there is no integer in the following interval:
\begin{equation}
    \label{eq:interval_I}
    I=\left[\frac{\kappa-1}{\Omega}, \frac{\kappa+1}{\Omega}\right].
\end{equation}

The interval $I$ is centered around $\kappa/\Omega$ and has a width $2/\Omega$. If the width of $I$ is greater than or equal to 1 (implying $\Omega<2$), there must always be an integer in $I$ for all values of $\kappa$. This means that there is at least one term of type \eqref{eq:fact1} in the decay rate sum \eqref{eq:helix_eigenvalue_real_part}, so no states are trapped \footnote{If the values of $\kappa$ and $\Omega$ are such that the argument of the Bessel function is a root for each term in the sum, that state will be trapped. These values of $\kappa$, however, form a set of measure 0, while the other condition discussed, namely $\Omega>2$, forms a set of measure 1.}. This is shown in the plots in Fig.~\ref{fig:infinite_helix} with $\Omega<2$: no regions of trapped states are visible. If $\Omega\geq2$, then there will be trapped states. Specifically, any eigenstate with $\kappa$ inside the following interval $T$ will be a trapped state: 
\begin{equation}
    T=\left[j\,\Omega+1,(j\!+\!1)\Omega\!-\!1\right]\,\,\forall \, j\in[0,1,...,\infty),\, \Omega\geq 2
    \label{eq:trapped_states_condition}
\end{equation}
We refer to these regions of contiguous trapped states as ``measure one sets of trapped states," as opposed to trapped states that occur only at one specific value (or finitely many values) of $\kappa$. If $\Omega=2$, $T=[2j+1,2j+1]$, which means that trapped states will appear at the odd integers. This is shown in the $\Omega=2$ plot of Fig.~\ref{fig:infinite_helix}. For $\Omega=3$, $T=[3j+1,3j+2]$. This means trapped states should appear when $\kappa$ is in $[1,2]$, $[4,5]$, and so on. This is indeed seen in the plots of Fig.~\ref{fig:infinite_helix} with $\Omega=3$. For $\Omega=5$, $T=[5j+1,5j+4]$, so trapped states should appear when $\kappa$ is in $[1,4]$, $[6,9]$, and so on. The first of these trapped ranges is in the $\Omega=5$ plot of Fig.~\ref{fig:infinite_helix}. For $\Omega=10$, $T=[10j+1,10j+9]$, so the first of the trapped ranges is $\kappa\in[1,9]$, which extends off to the right of the $\Omega=10$ plot.

The periodicity of this domain means that we can define a notion of the ``fraction" of states that are trapped. Consider the start of one of the trapped $\kappa$ domains at $j\,\Omega+1$. The next trapped domain will start at $(j+1)\Omega+1$. So, the length between the start of two trapped domains is $\Omega$. The length of one trapped domain is $\Omega-2$. So, the fraction of the total interval that is trapped is $(\Omega-2)/\Omega$.

\subsection{Superradiant and subradiant states}
To see which states are superradiant ($\Gamma/\gamma>1$) and which are subradiant ($\Gamma/\gamma<1$), we can use the expression for the decay rates in Eq.~\eqref{eq:helix_eigenvalue_real_part}. Solving for $\Gamma/\gamma$, we obtain
\begin{equation}
    \frac{\Gamma}{\gamma}=\frac{2\pi n_0}{k_0}\sum_{m=m_\text{min}}^{m_\text{max}}J_m^2\left(\sqrt{1-(\kappa-m\Omega)^2}\,r\right)
    \label{eq:gamma_over_gamma_helix}
\end{equation}
where $m_{\min}$ and $m_{\max}$ are defined in Eq.~\eqref{eq:m_min_and_m_max}. The superradiant states occur at any values of $\kappa$ for which $\Gamma/\gamma>1$. The prefactor $\frac{2\pi n_0}{k_0}$ scales up all the plots in Fig.~\ref{fig:infinite_helix} without changing the functional form. The simplest case is if $\Omega\geq2$, since then $[m_{\min},m_{\max}]$ contains at most one integer. In this case, plugging the maximal decay rate state $\kappa=1$ into Eq.~\eqref{eq:gamma_over_gamma_helix} will reduce the sum to just the $m=0$ term (and the $m=1$ term if $\Omega=2$, but this term is 0), making $\Gamma/\gamma=2\pi n_0/k_0$. So, for the $\kappa=1$ state to be superradiant, we must have $2\pi n_0/k_0>1$. If $\frac{2\pi n_0}{k_0}<1$, then all states will be subradiant.

For the case $\Omega<2$, the value of the sum \eqref{eq:gamma_over_gamma_helix} will be the sum of multiple terms, because $[m_\text{min},m_\text{max}]$ will have multiple integers. The state at $\kappa=1$ will be superradiant simply if the prefactor $\frac{2\pi n_0}{k_0}$ takes a value such that the expression \eqref{eq:gamma_over_gamma_helix} is greater than 1, and subradiant otherwise.

\subsection{Thermally averaged decay rates}
\begin{figure}
    \centering
    \includegraphics[width=0.6\linewidth]{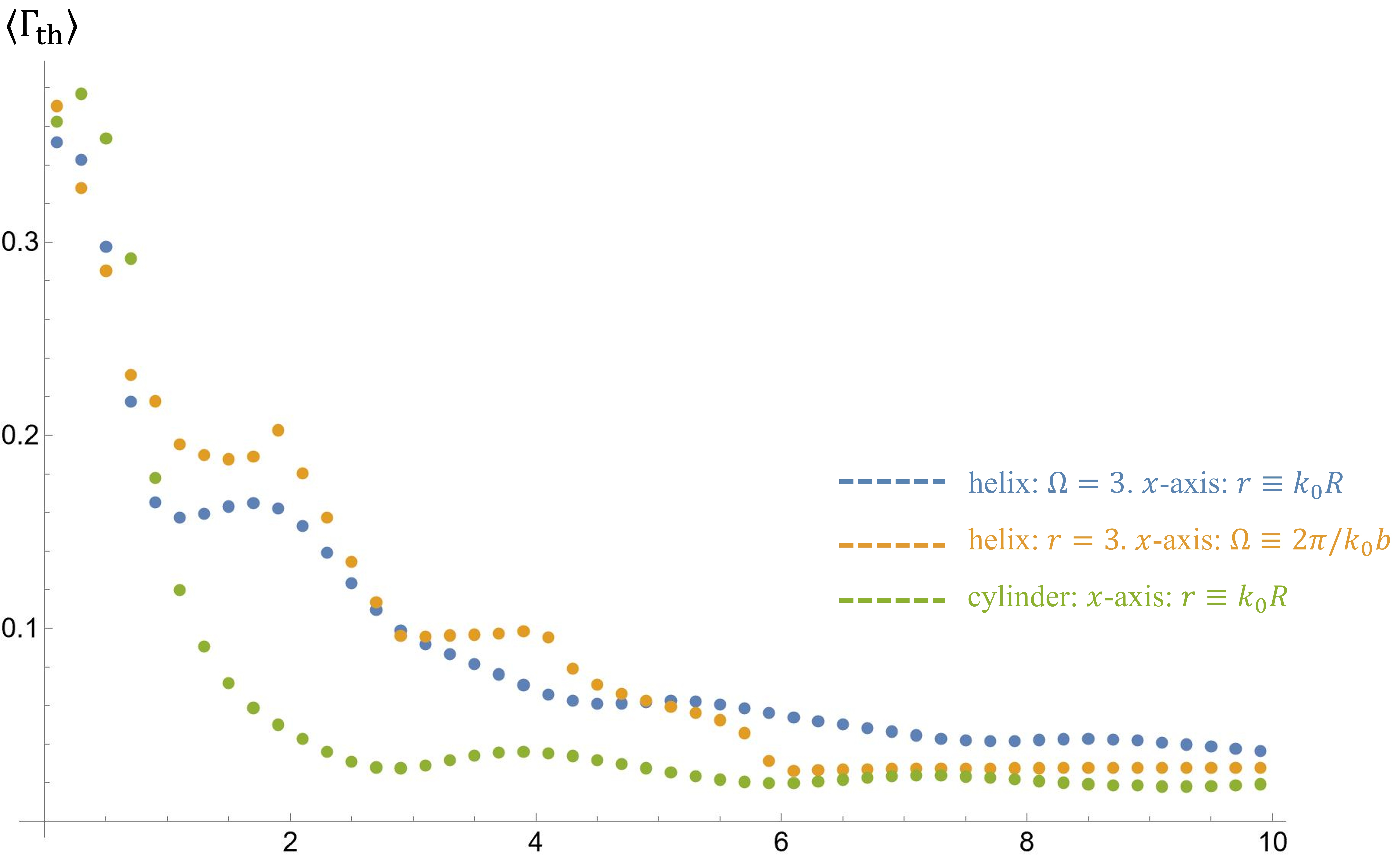}
    \caption{\label{tab:thermal_Gamma}\textbf{The thermally averaged collective decay rate $\Gammath$ of an infinite helix of quantum emitters is consistently larger than that of an infinite cylinder, due to more eigenstates in the helices having high decay rates and large-magnitude, negative collective Lamb shifts.}  The vertical axis value of each point represents the value of the integral in Eq.~\eqref{eq:thermal_Gamma_modified} for the parameters specified by the $x$-axis and the legend, where we have set $\beta=1$. The dimensionless values $r\equiv k_0 R$ and $\Omega\equiv 2\pi/k_0 b$ are used on the $x$-axis, where $k_0$ is the excitation wavenumber, $b$ is the helical pitch, and $R$ is the radius. The blue (orange) points are for the infinite continuous helix, where $\Omega$ ($r$) is kept constant at $3$, and $r$ ($\Omega$) is given by the $x$-axis values. The green points are for the infinite cylinder solution from Ref. \cite{Svidzinsky2016}. For the cylinder there is no $\Omega$ value, so $r$ is varied on the $x$-axis. The integral in Eq.~\eqref{eq:thermal_Gamma_modified} was approximated by discretizing $\kappa$ into intervals of length $\Delta\kappa=0.01$, such that $\kappa_i\in[0,0.01,\,...\,,4.99,5]$, and turning the integral into a sum. }
\end{figure}

It is possible to define a thermally averaged collective decay rate $\Gammath$ as the following:
\begin{equation}
    \Gammath\equiv \frac{\int d\kappa \,\,\Gamma(\kappa) \,e^{-\beta E(\kappa)}}{\int d\kappa\, e^{-\beta E(\kappa)}}
    \label{eq:thermal_Gamma_continuous}
\end{equation}
where $\beta\equiv (k_BT)^{-1}$. This is the continuous version of the thermally averaged decay rate used in steady-state fluorescence quantum yield measurements of superradiance \cite{Babcock2024}, which contains a sum over discrete eigenvalues rather than an integral. 

However, the divergence of $E(\kappa)$ to $-\infty$ at $\kappa=1$ makes the integrals in Eq.~\eqref{eq:thermal_Gamma_continuous} intractable, so we rescale the collective Lamb shifts $E(\kappa)$. A collective Lamb shift of $-\infty$ for the $\kappa=1$ state means physically that the $\kappa=1$ state has the lowest energy in the eigenspectrum, which we can set to $0$, giving a weight in the thermal Gibbs ensemble of $\exp(0)=1$. The state with the highest collective Lamb shift should then correspondingly have a weight of $\exp(-\infty)=0$. So, we can modify the expression in Eq.~\eqref{eq:thermal_Gamma_continuous} by considering a function $f$ of the collective Lamb shifts:
\begin{equation}
    \Gammath\equiv \frac{\int d\kappa \,\,\Gamma(\kappa) \,f(E(\kappa))}{\int d\kappa\, f(E(\kappa))},
     \label{eq:thermal_Gamma_modified_f}
\end{equation}
such that the weight in the thermal Gibbs ensemble of the $E=-\infty$ state is $1$ and the corresponding weight of the maximum $E$ state is $0$. A straightforward mapping function that preserves the weighting scheme of the Boltzmann distribution is
\begin{equation}
    \Gammath\equiv \frac{\int d\kappa \,\,\Gamma(\kappa) \left[1-c\,e^{\beta E(\kappa)}\right]}{\int d\kappa \left[1-c\,e^{\beta E(\kappa)}\right]},
    \label{eq:thermal_Gamma_modified}
\end{equation}
where $c = \exp(-\beta E^{\max}(\kappa))$ represents the Boltzmann weight for the finite maximum value of $E$ across all $\kappa$ in each eigenspectrum.

Using this function $f(E(\kappa))$ in Eqs.~\eqref{eq:thermal_Gamma_modified_f} and \eqref{eq:thermal_Gamma_modified}, the thermally averaged decay rates $\Gammath$ can be computed. A plot of $\Gammath$ for many $r$ and $\Omega$ values is given in Fig.~\ref{tab:thermal_Gamma}, along with the $\Gammath$ values of the infinite continuous cylinder solution from Ref.~\cite{Svidzinsky2016} for comparison. The values of $\Gammath$ for the helices are greater than those of the infinite cylinder across all parameter regimes (except for when $\Omega$ or $r$ goes to 0, at which they all converge). Looking at Fig.~\ref{fig:infinite_helix}, we can see why. For small $\Omega$ or $r$ (the leftmost plots in Fig.~\ref{fig:infinite_helix}), there are many large decay rate values when $\kappa<1$, as well as regions of non-trapped states when $\kappa>1$. Almost all states have negative values of the collective Lamb shift, with the few exceptions showing in the four rightmost panels of Fig.~\ref{fig:infinite_helix}. These results highlight the possibilities for engineering helical architectures in thermal environments to maximize radiative decay rates, quantum yields, and collective error correction via dissipation to the electromagnetic field, by tuning just the geometric free parameters.

Note that the values of $\kappa$ for which trapped states occur, given in Eq.~\eqref{eq:trapped_states_condition}, do not depend on the radius $k_0R$ of the helix. The radius does not affect which regions exhibit only trapped states for $\kappa>1$: rather, it affects how much the collective decay rates (and Lamb shifts) of the non-trapped states oscillate with $\kappa$, per the top row of panels in Fig.~\ref{fig:infinite_helix}.

\section{\label{sect:comparison}Comparisons of infinite cylinder, helix, and lines}
\subsection{\label{sect:cylinder_helix_comparison}Comparison of infinite continuous cylinder and helix}
Since the helix lies on the surface of the infinite continuous cylinder surface studied in \cite{Svidzinsky2016}, we can expect that there will be correspondences between the two solutions, and indeed there are. Consider the infinite cylinder eigenvalues derived in \cite{Svidzinsky2016}, re-written in Eqs. \eqref{eq:svidzinsky_cylinder_Gammas} and \eqref{eq:svidzinsky_cylinder_Energies}. In the axially symmetric case where $n=0$, the eigenvalues become
\begin{equation}
    \mathcal{E} = \frac{\pi \gamma n_0}{k_0}J_0\left(\sqrt{k_0^2-k_z^2}R\right) H_0^{(1)}\left(\sqrt{k_0^2-k_z^2}R\right)\nonumber,
\end{equation}
and the eigenfunctions become
\begin{equation}
    \beta(\theta,z)=\beta(z)=e^{ik_z z}.\nonumber
\end{equation}
These eigenfunctions match with those of the infinite helix. The real part of the eigenvalues (i.e., the decay rates) also match in certain parameter regimes. If the interval $I$ in Eq.~\eqref{eq:interval_I} contains \textit{only} $0$, then the only term in the sum for the real part of \eqref{eq:helix_eigenvalue} will be the $m=0$ term. The interval $I$ contains $0$ only when $\Omega>2$ and $\kappa<1$, or when $1<\Omega<2$ and $|\kappa|<\Omega-1$. In other words, if $\Omega>1$, there are certain values of $\kappa$ for which the interval $I$ contains only 0. For these $\Omega$ and $\kappa$ values, the decay rate \eqref{eq:helix_eigenvalue_real_part} becomes
\begin{equation}
    \Gamma = \frac{\pi\gamma n_0}{k_0}J_0^2\left(\sqrt{1-\kappa^2}\,r\right)\nonumber,
\end{equation}
which is the same as the real part of the infinite cylinder eigenvalue. 

Physically, the condition $\Omega>1$ is equivalent to the condition $\lambda_0/b>1$ by definition ($\Omega\equiv2\pi/k_0 b$). So, the equivalence between the infinite cylinder and helix only emerges when the pitch $b$ becomes smaller than the excitation wavelength $\lambda_0$. In this ``long-wavelength'' regime, the helical turns fall outside resolution and appear grossly like the surface of a cylinder.

\subsection{\label{sect:estimates_biostructures}Comparison of infinite continuous line and helix}
In this section, the infinite continuous helix and infinite continuous line solutions derived in this paper are compared. There are two limits in which the helix approaches a line: the $r\rightarrow 0$ limit, and the $\Omega\rightarrow 0$ limit. In both of these limits, the helix decay rates become equal to the line decay rates. In the formal limit $\Omega\rightarrow 0$, the decay rates from Eq.~\eqref{eq:helix_eigenvalue_real_part} become
\begin{align}
    \frac{k_0}{2\pi \gamma n_0}\Gamma \,\,&=\lim_{\Omega\rightarrow 0} \sum_{m=-\infty}^{\infty} J_m^2\left(\sqrt{1-(\kappa-m\Omega)^2}\,r\right)\nonumber\\
    &=\sum_{m=-\infty}^{\infty} J_m^2\left(\sqrt{1-\kappa^2}\,r\right)\label{eq:Omega_goesto_0_line2}\\
    &=\begin{cases}
        1\,\,\,\,\,\,\, &|\kappa|\leq1\\
        0 &|\kappa|>1.
    \end{cases}\label{eq:Omega_goesto_0_line3}
\end{align}
The step from Eq.~\eqref{eq:Omega_goesto_0_line2} to Eq.~\eqref{eq:Omega_goesto_0_line3} follows from the identity $\sum_{m=-\infty}^{\infty}J_m^2(x)=1$ for any real $x$. For any imaginary $x$, by Eq.~\eqref{eq:fact4}, the real part of $J_m(x)$ is 0. If $|\kappa|<1$, the argument is real for all terms in the sum and the above identity makes the sum $1$, and if $|\kappa|>1$, the argument is imaginary for all terms, making the sum $0$. This is the infinite line decay rate expression from Eq.~\eqref{eq:line_gamma}.

For the formal limit $r\rightarrow 0$, we have
\begin{align}
    \frac{k_0}{2\pi \gamma n_0}\Gamma \,\,&=\lim_{r\rightarrow 0} \sum_{m=m_\text{min}}^{m_\text{max}} J_m^2\left(\sqrt{1-(\kappa-m\Omega)^2}\,r\right)\nonumber\\
    &=\begin{cases}
        J_0^2(x)\,\,\,\,\,\, &|\kappa|\leq1\\
        0 &|\kappa|>1
    \end{cases}\label{eq:r_goesto_0_line2}\\
    &=\begin{cases}
        1\,\,\,\,|\kappa|\leq1\\
        0\,\,\,\,|\kappa|>1.
    \end{cases}\label{eq:r_goesto_0_line3}
\end{align}
As $r\rightarrow 0$, the only argument of the Bessel function that matters will be $0$. For the argument being $0$, the only Bessel function order $m$ that has a nonzero contribution is $m=0$. The interval $[m_{\min},m_{\max}]$ only contains $0$ if $|\kappa|<1$. So, if $\kappa>1$, all the terms will be 0, and if $|\kappa|<1$, the only term is $J_0^2(0)$. Eq.~\eqref{eq:r_goesto_0_line2} follows from this, and from that follows Eq.~\eqref{eq:r_goesto_0_line3}, which is the solution \eqref{eq:line_gamma} for the infinite continuous line decay rates.

Aside from these limits in which the helix collapses into a line, the solutions of the line and the helix are quite different. The extra complexity of the helical solution arises from the two extra geometric parameters that come with the helix: the helical pitch $b$ and the radius $r$. In fact, the helix has more geometric parameters than the topologically two-dimensional infinite cylinder, which is described with just $r$. This gives rise to novel features of the helix that are not present in the infinite cylinder, such as the existence of non-trapped states when $\kappa>1$.

\subsection{\label{sect:discrete_line_vs_continuous_line}Comparison between discrete and continuous models of the infinite line}

\begin{figure}
    \centering
    \includegraphics[width=0.5\linewidth]{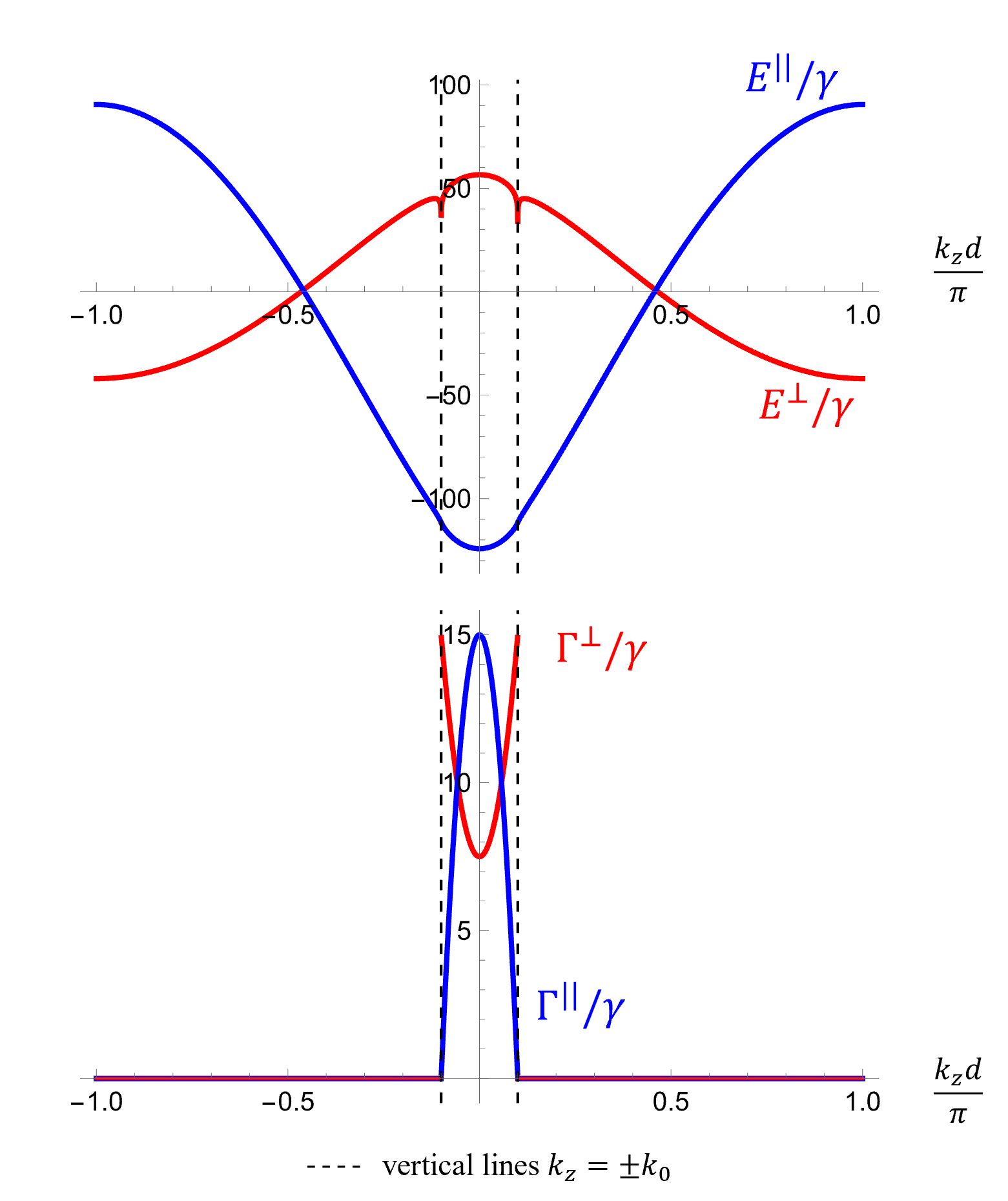}
    \caption{\textbf{Collective Lamb shifts $E^\|$ and $E^\perp$ and radiative decay rates $\Gamma^\|$ and $\Gamma^\perp$ for a discrete line of quantum emitters modeled as transition dipole vectors in the limit of zero emitter spacing are distinct from those for a continuous line of scalar emitters.} The collective Lamb shifts and decay rates from Ref.~\cite{AAG2017}, re-written here in Eqs. \eqref{eq:E_par}-\eqref{eq:AAG_Gamma_perp}, are plotted as a function of $k_z d/\pi$ for the infinite discrete line. Since the functions diverge when $d = 0$, the value $d=0.05$ was used for plotting purposes. The black vertical dashed lines indicate where $k_z=\pm k_0$. Note that there is a vertical asymptote for $E^{\perp}$ at $k_z=\pm k_0$, because at these values $E^{\perp}$ diverges to $-\infty$. These plots should be compared to Fig.~\ref{fig:infinite_line_vs_kz_k0}, given by Eq.~\eqref{eq:line_eigvals}; though there are similarities, it can be seen that neither $E^\|$ nor $E^\perp$ converge to the collective Lamb shift of the infinite continuous line in the limit $d\rightarrow 0$. The same applies for $\Gamma^\|$ and $\Gamma^\perp$. This difference occurs because the interaction terms in the Hamiltonian from Ref. \cite{AAG2017} contain terms proportional to $1/r^2$ and $1/r^3$, which are not present in the Hamiltonian used in Fig. \ref{fig:infinite_line_vs_kz_k0}, and the emitters in the infinite continuous line from Fig. \ref{fig:infinite_line_vs_kz_k0} are scalar objects, with their polarization effects neglected.}
    \label{fig:AAG_2017_limit_d_goesto0}
\end{figure}

An interesting comparison can be made between a discrete, infinite line of quantum emitters, which was studied in Ref.~\cite{AAG2017}, and the continuous, infinite line of emitters studied in this work. The collective Lamb shifts $E$ calculated in Ref.~\cite{AAG2017} are 
\newcommand{\Li}{\text{Li}}
\begin{align}
    E^{\|}&= -\frac{3\gamma}{2(k_0d)^3}\Re\Big[\Li_3(e^{i(k_0+k_z)d})+\Li_3(e^{i(k_0-k_z)d})\nonumber\\
    &-ik_0d\left(\Li_2(e^{i(k_0+k_z)d})+\Li_2(e^{i(k_0-k_z)d})\right)\Big]
    \label{eq:E_par}
\end{align}
\begin{align}
    E^{\perp}&= \frac{3\gamma}{4(k_0d)^3}\Re\Big[\Li_3(e^{i(k_0+k_z)d})+\Li_3(e^{i(k_0-k_z)d})\nonumber\\
    &-ik_0d\left(\Li_2(e^{i(k_0+k_z)d})+\Li_2(e^{i(k_0-k_z)d})\right)\nonumber\\
    &+k_0^2d^2\left(\ln(1-e^{i(k_0+k_z)d})+\ln(1-e^{i(k_0-k_z)d})\right)\Big]
    \label{eq:E_perp}
\end{align}
and the decay rates are
\begin{align}
    \Gamma^{\|}&=\frac{3\gamma\pi}{2k_0d}\sum_{g_z=\left\lceil\frac{-k_0-k_z}{2\pi}\right\rceil}^{\left\lfloor\frac{k_0-k_z}{2\pi}\right\rfloor}\left(1-\frac{(k_z+g_z)^2}{k_0^2}\right)
    \label{eq:AAG_Gamma_par}
\end{align}
\begin{align}
    \Gamma^{\perp}&=\frac{3\gamma\pi}{2k_0d}\sum_{g_z=\left\lceil\frac{-k_0-k_z}{2\pi}\right\rceil}^{\left\lfloor\frac{k_0-k_z}{2\pi}\right\rfloor}\left(1+\frac{(k_z+g_z)^2}{k_0^2}\right).
    \label{eq:AAG_Gamma_perp}
\end{align}
The superscripts $\perp$ and $\|$ denote the orientation of the transition dipole vectors of the quantum emitters: $\|$ means that the vectors are oriented parallel to the line of emitters, and $\perp$ means that they are perpendicular to the line of emitters. In the approach of Ref.~\cite{Svidzinsky2016}, which is what we use to obtain the infinite continuous line and helix results, the polarization information is neglected (i.e., there are no transition dipole vectors). The qualitative structure of our continuous line solution in Eqs.~\eqref{eq:line_gamma}-\eqref{eq:line_E} and the discrete line solution in Eqs.~\eqref{eq:E_par}-\eqref{eq:AAG_Gamma_perp} are similar in some ways. For the decay rates, in Eqs.~\eqref{eq:line_gamma}, \eqref{eq:AAG_Gamma_par}, and \eqref{eq:AAG_Gamma_perp}, there are trapped states for any $k_z$ value that has magnitude greater than $k_0$ ($|\kappa|>1$). For the collective Lamb shifts, both Eq.~\eqref{eq:line_E} and Eq.~\eqref{eq:E_perp} have divergences to $-\infty$ when $k_z=\pm k_0$ ($\kappa=\pm 1$). There is also a concave-down portion of the collective Lamb shift function that is seen from Ref. \cite{AAG2017} in their red curve of Fig. 1b and of the decay rates in their blue curve of Fig. 1c, which is reproduced with $d/\lambda_0=0.05$ in our Fig.~\ref{fig:AAG_2017_limit_d_goesto0}. Their results are similar to the qualitative collective Lamb shift and decay rate curve features in our Fig.~\ref{fig:infinite_line_vs_kz_k0}.

One may initially expect that the correspondence is more than just qualitative and that the solutions become exactly the same when the emitter spacing $d$ in the discrete case is taken to $0$. However, this is not the case, as seen in Fig.~\ref{fig:AAG_2017_limit_d_goesto0}. Looking at the decay rates, for both the transverse and longitudinal solutions from Ref.~\cite{AAG2017}, as $d\rightarrow0$, the maximum value of the functions approach $+\infty$, and the parabolic structure for $|k_z|<k_0$ shown in Fig. 1b and 1c in their work remains (it does not flatten out).

There are multiple reasons why the two solutions do not converge in the limit $d\rightarrow 0$. First, the analytical solution in Ref.~\cite{AAG2017} was derived using a discrete Fourier transform. To properly account for the limit $d\rightarrow 0$ and avoiding divergences, the discrete Fourier transform would have to be changed to a continuous Fourier transform. The resulting collective Lamb shifts and decay rates would involve continuous Fourier transforms of three terms, proportional to $\exp(ik_0(z-z'))/(z-z')$, $\exp(ik_0(z-z'))/(z-z')^2$, and $\exp(ik_0(z-z'))/(z-z')^3$. The Fourier transform of the first term is exactly what is used in our infinite continuous line derivation (see Eq.~\eqref{eq:FT_of_K}). The Fourier transforms of the other two terms would serve as corrections to the infinite continuous line eigenvalues \eqref{eq:line_gamma} and \eqref{eq:line_E}.

\textcolor{black}{Why does our approach, using the eigenvalue equation \eqref{eq:eigenvalue-eq}, only have the $1/r$ interaction term, while the approach from Ref.~\cite{AAG2017} has additional interaction terms proportional to $1/r^2$ and $1/r^3$?} In the approach of Refs.~\cite{AAG2017, Patwa2024, Babcock2024}, each two-level system is mathematically treated as a dipole with dipole moment operator $\hat{\mathbf{p}}_j=\bm{\wp}_j^{*}|e_j\rangle\langle g_j|+\bm{\wp}_j|g_j\rangle\langle e_j|$ where $|e_j\rangle\langle g_j|$ ($|g_j\rangle\langle e_j|$) represents the raising (lowering) operator of the two-level system and the vector $\bm{\wp}_j\equiv q\langle e_j|\hat{\mathbf{r}}|g_j\rangle$ is the transition dipole vector, with $q$ being the charge difference between the excited and ground states for each emitter. This means that the interaction Hamiltonian contains a term with the scalar product $\bm{\wp}_j\cdot \mathbf{\hat{E}(\mathbf{r})}$, where $\mathbf{\hat{E}(\mathbf{r})}$ is the electric field vector operator. This interaction Hamiltonian describes the interaction between a collection of two-level systems and the electromagnetic field. A Lindblad equation, which models the two-level system network as an open quantum system, is then obtained in Ref.~\cite{AAG2017} by tracing out the electromagnetic degrees of freedom, which involves integrating $\bm{\wp}_j\cdot \mathbf{\hat{E}(\hat{\mathbf{r}})}$ over the angular coordinates $\theta$ and $\phi$. From this integration, the $1/r$, $1/r^2$, and $1/r^3$ terms emerge. On the other hand, in Ref.~\cite{Svidzinsky2016}, the two-level systems are considered as scalars, rather than transition dipole vectors. So, the scalar product $\bm{\wp}_j\cdot \mathbf{\hat{E}(\hat{\mathbf{r}})}$ is reduced to the scalar multiplication $\wp|\hat{\mathbf{E}}(\hat{\mathbf{r}})|$. The Schr\"{o}dinger equation with the interaction Hamiltonian including the term $\wp|\hat{\mathbf{E}}(\hat{\mathbf{r}})|$ is then solved for the probability amplitudes for single-excitation Fock states, and from that the $1/r$ term emerges in the eigenvalue equation, Eq.~\eqref{eq:eigenvalue-eq}.

Note that the interaction terms proportional to $1/r$, $1/r^2$, and $1/r^3$---obtained using the open quantum systems approach described above---exactly match the interaction terms obtained from the canonical electromagnetic Green's function for a network of \textit{classical} point dipoles (replacing each discrete transition dipole with a Hertzian one) including all near- to far-zone contributions. This is an intriguing correspondence, and it is exploited in Ref.~\cite{AAG2017}, instead of presenting the open quantum systems derivation described above, which is detailed further in Appendix H of Ref.~\cite{mattiotti2022}.

\section{Estimates of superradiance and subradiance in realistic protein fibers}
\begin{figure*}
    \centering
    \,\\\,\\
    \textbf{a)}\,\,\,\,\,\,\,\,\,\,\,\,\,\,\,\,\,\,\,\,\,\,\,\,\,\,\,\,\,\,\,\,\,\,\,\,\,\,\,\,\,\,\,\,\,\,\,\,\,\,\,\,\,\,\,\,\,\,\,\,\,\,\,\,\,\,\,\,\,\,\,\textbf{b)}\,\,\,\,\,\,\,\,\,\,\,\,\,\,\,\,\,\,\,\,\,\,\,\,\,\,\,\,\,\,\,\,\,\,\,\,\,\,\,\,\,\,\,\,\,\,\,\,\,\,\,\,\,\,\,\,\,\,\,\,\,\,\,\,\,\,\,\,\,\,\,\,\textbf{c)}\,\,\,\,\,\,\,\,\,\,\,\,\,\,\,\,\,\,\,\,\,\,\,\,\,\,\,\,\,\,\,\,\,\,\,\,\,\,\,\,\,\,\,\,\,\,\,\,\,\,\,\,\,\,\,\,\,\,\,\,\,\,\,\,\,\,\,\,\,\,\,\textbf{d)}\,\,\,\,\,\,\,\,\,\,\,\,\,\,\,\,\,\,\,\,\,\,\,\,\,\,\,\,\,\,\,\,\,\,\,\,\,\,\,\,\,\,\,\,\,\,\,\,\,\,\,\,\,\,\,\,\,\,\,\,\,\,\,\,\,\,\,\,\,\,\,\,\\
    
    \includegraphics[width=0.24\textwidth]{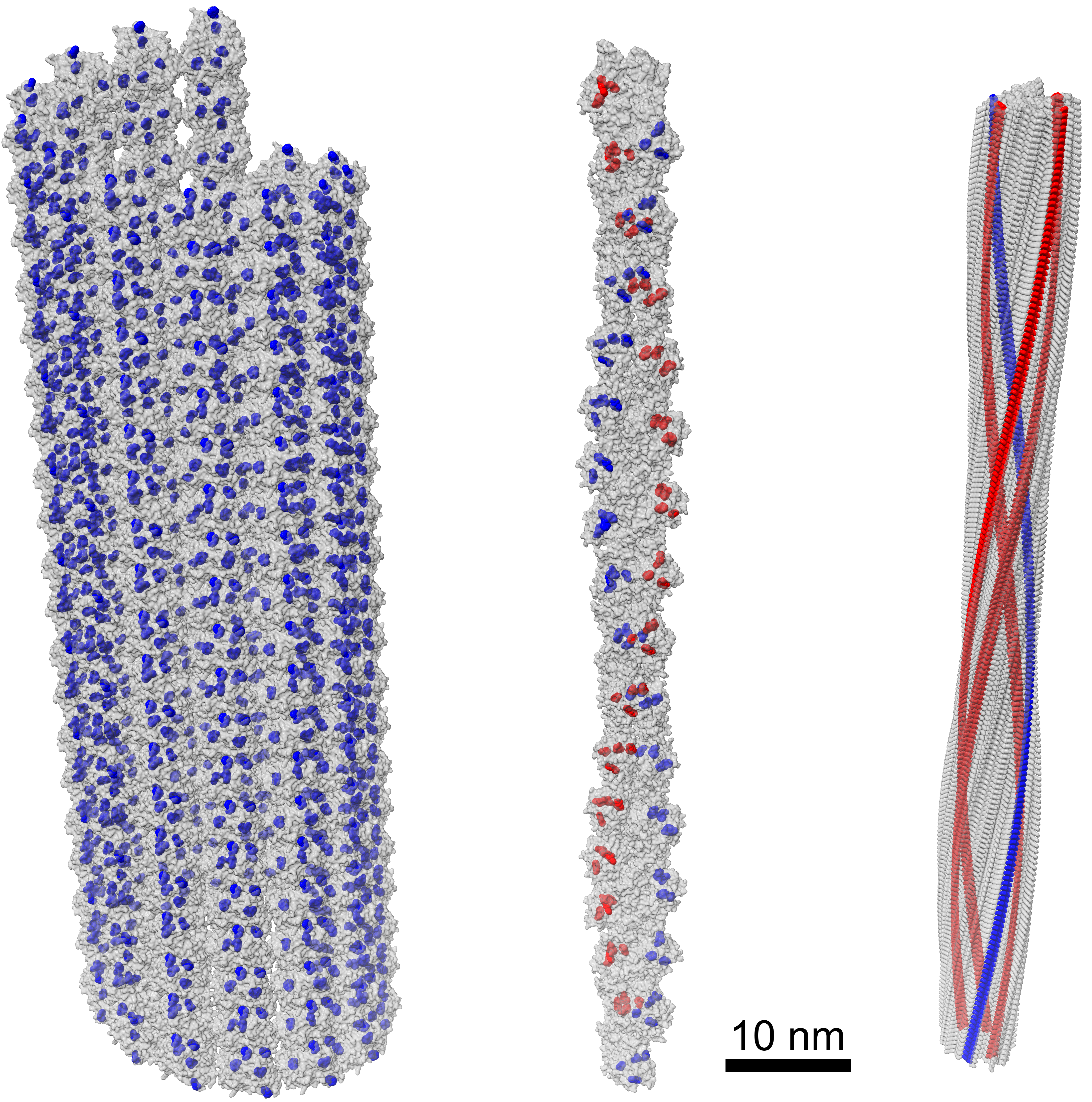}
    \includegraphics[width=0.24\textwidth]{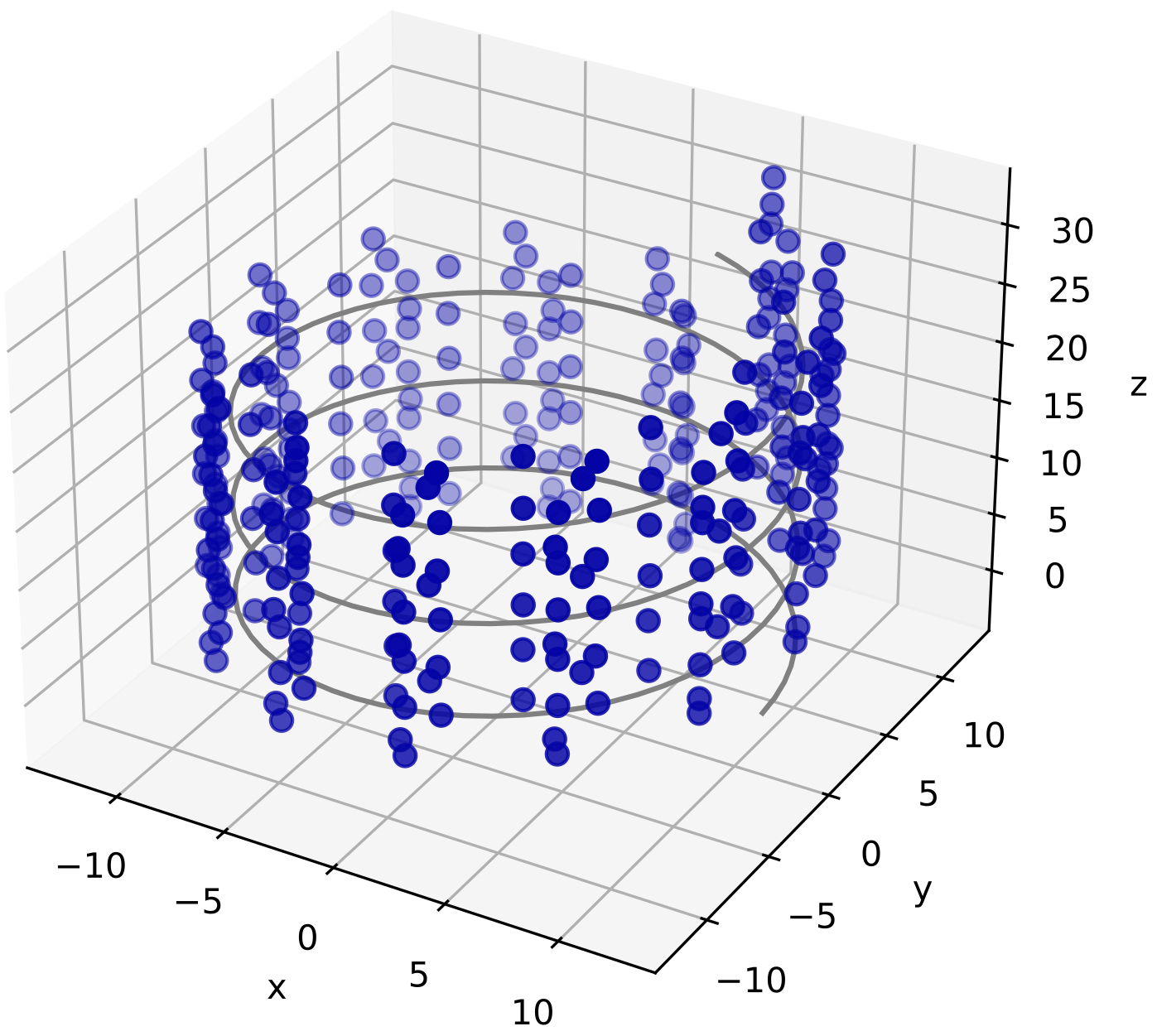}
    \includegraphics[width=0.24\textwidth]{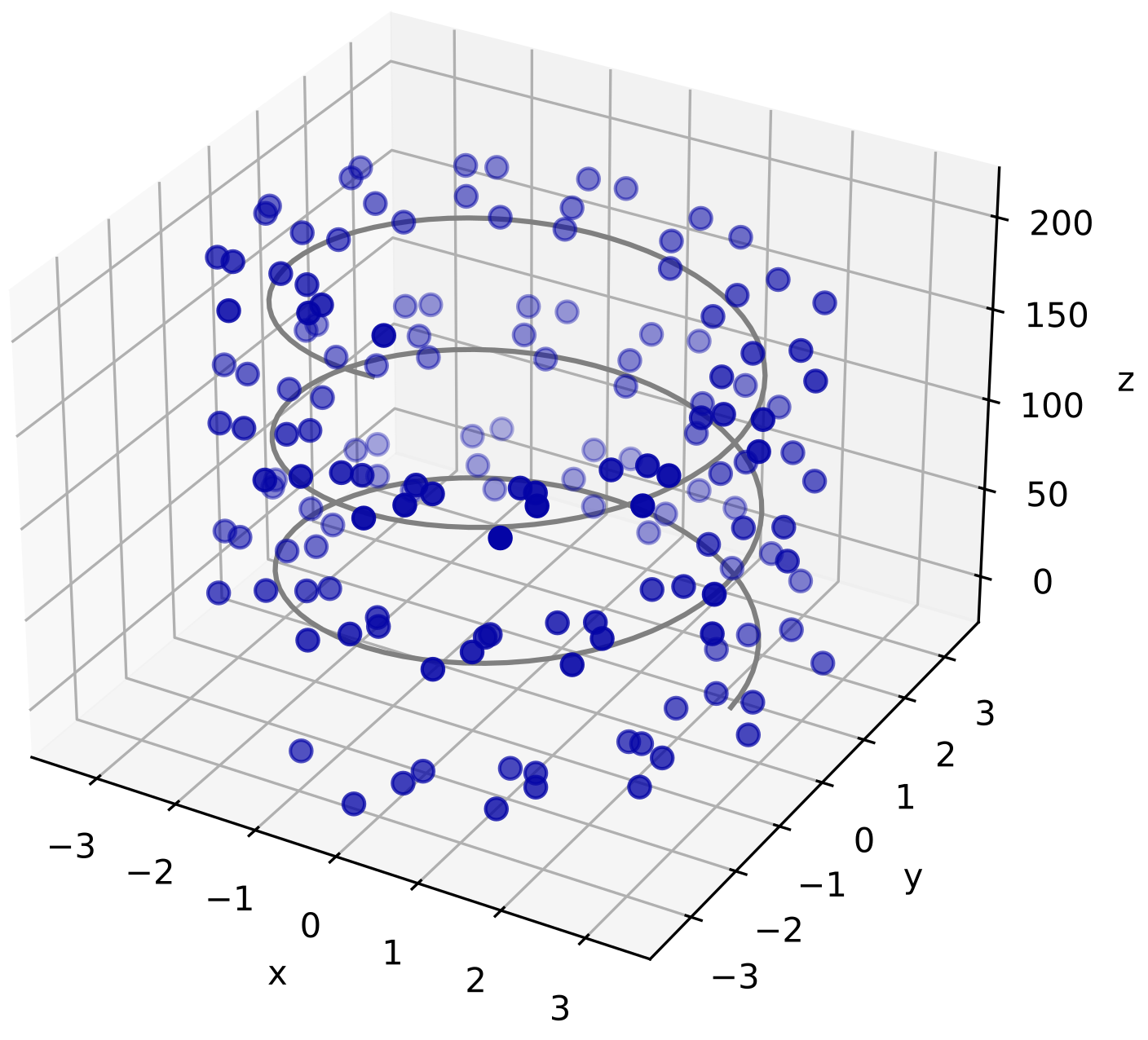}
    \includegraphics[width=0.24\textwidth]{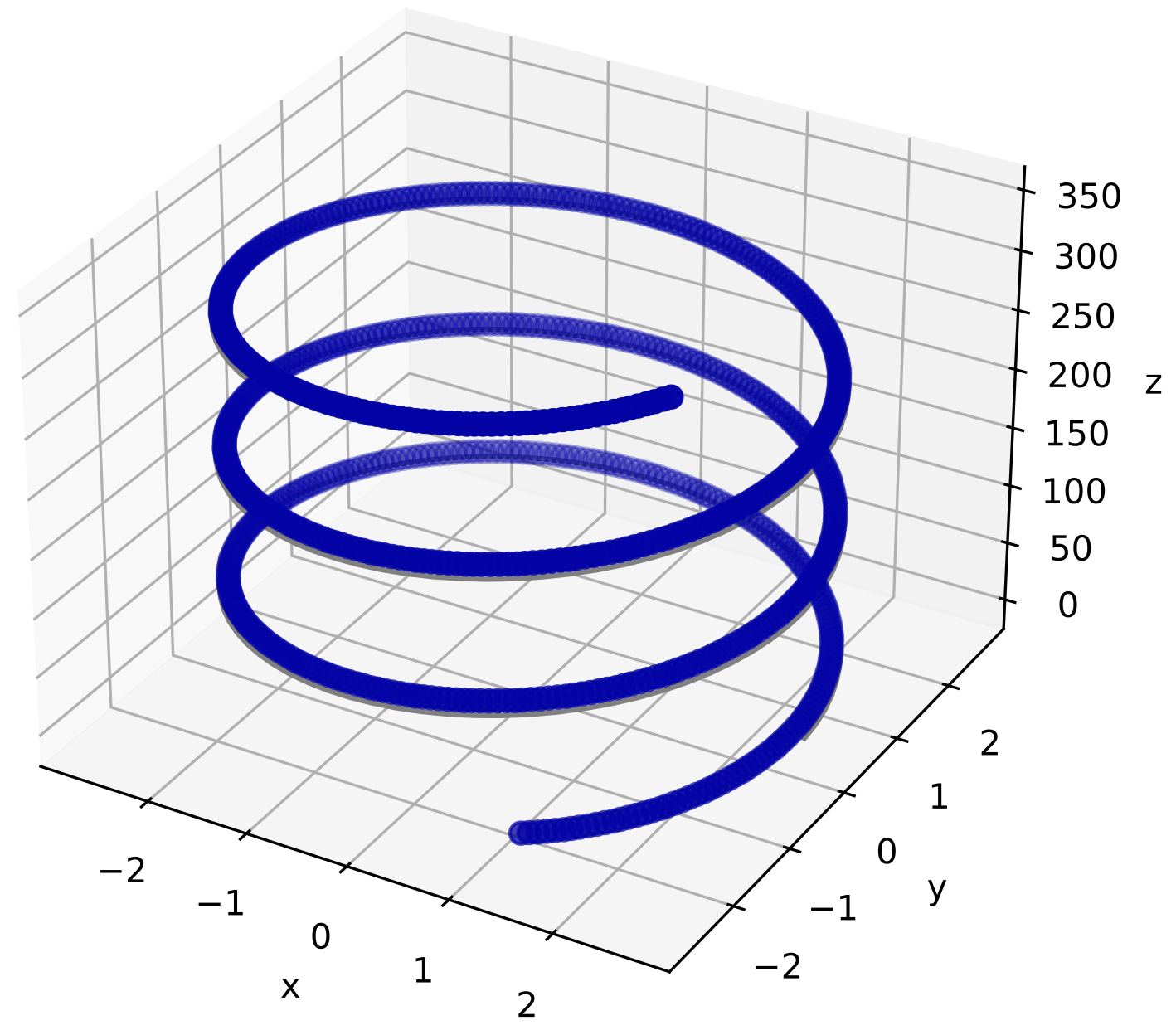}\\\textbf{e)}\\
    \includegraphics[width=\textwidth]{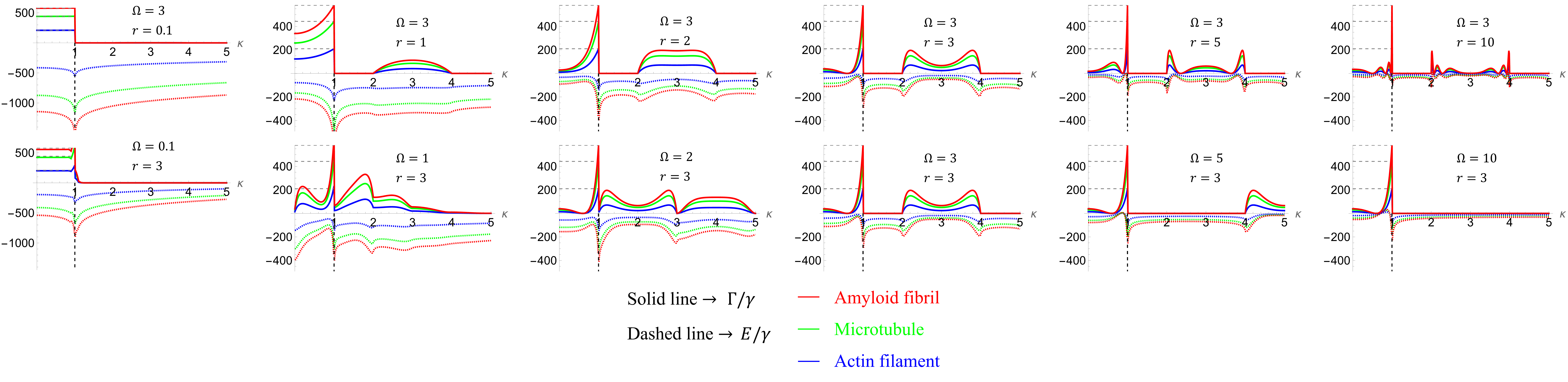}
    \caption{\textbf{The infinite continuous helix provides a good approximation of the photophysics of densely spaced networks of molecular quantum emitters (tryptophans) in protein fibers.} Panel \textbf{a)} shows, from left to right, a microtubule, an actin filament, and an amyloid fibril with the tryptophan amino acids highlighted in blue and red. Only the blue tryptophans are used to make the helical approximation of each structure. Panels \textbf{b)}, \textbf{c)}, and \textbf{d)} show only the blue tryptophans in the microtubule, actin filament, and amyloid fibril, respectively, from panel \textbf{a)}. The best helical approximation of each tryptophan helix in panels \textbf{b)}, \textbf{c)}, and \textbf{d)} is overlaid in gray. The axes in panels \textbf{b)}, \textbf{c)}, and \textbf{d)} are all in units of nm. Panel \textbf{e)} shows the eigenspectrum of the infinite helix approximations of each structure for varying parameters $\Omega$ and $r$, but using the actual line density values $n_0$ from each protein fiber. The advantage of this data representation is that $\Gamma/\gamma$ can directly indicate which states are superradiant ($\Gamma/\gamma>1$) and subradiant ($\Gamma/\gamma<1$), rather than only trapped and non-trapped as in Fig.~\ref{fig:infinite_helix}.}
    \label{fig:trp_positions}
\end{figure*}
\begin{table*}
    \begin{tabularx}{\textwidth}{|p{0.35\textwidth}|p{0.15\textwidth}|p{0.2\textwidth}|X|}
    \hline
    & Microtubule & Actin filament & Amyloid fibril \\
    \hline
    Trp network avg. $R$ (nm) & 11.2 & 2.64 & 2.71\\
    Trp network avg. $b$ (nm) & 7.8 & 73.1 & 112.3\\
    Trp network avg. $n_0$ (nm$^{-1}$) & 1.58 & 0.75 & 2.07\\
    Trp network $\Omega=2\pi/k_0b$ & 35.9 & 3.83 & 2.49\\
    Trp network $r=k_0 R$ & 0.25 & 0.06 & 0.06\\
    \hline
   $\Gamma_{\text{max}}/\gamma$ analytical from helix & 442.4 & 210 & 579.6\\
    $\Gamma_{\text{max}}/\gamma$ numerical from protein & 573.5 & 9.8 (6.6, SH) & 368.7 (224.2, SH) \\
    \hline
    \% $\Gamma_j$ trapped from helix & 94.43 & 47.79 & 19.79\\
    \% $\Gamma_j$ subradiant from protein & 99.99 & 65.13 (59.03, SH) &98.14 (98.47, SH)\\
    \hline
    \end{tabularx}
    \caption{\label{tab:estimates_biostructures}\textbf{The eigenvalues of an infinite continuous helix of tryptophan (Trp) emitters provide an efficient order-of-magnitude estimate of the maximum decay rate and percentage of trapped states for protein fiber architectures, which exist natively in parameter regimes with high thermally averaged decay rates (small $\Omega$ and/or $r$).} Numerical simulations of protein fiber models from Ref.~\cite{Patwa2024} were used for comparison. The values in rows 1-5, 7, and 9 from the table are calculated from only the blue tryptophans in the protein models shown in Fig.~\ref{fig:trp_positions}. $R$ is the radius, $b$ is the helical pitch, $n_0$ is the line density, $\Omega$ is the dimensionless inverse pitch, and $r$ is the dimensionless radius (calculated from $b$ and $R$, respectively). The helix decay rates given in Eq.~\eqref{eq:helix_eigenvalue_real_part}, evaluated for the given table parameters, are used to calculate the maximum decay rates in row 6. The percentage of trapped states reported in row 8 was calculated using the formula $(\Omega-2)/\Omega$, which was derived in Section \ref{sec:trapped_states}. For the actin filament and the amyloid fibril, the values in parentheses are the numerically calculated predictions for only a single helix (SH) of emitters within the protein fiber: in other words, using only the blue tryptophans from Fig.~\ref{fig:trp_positions}\textbf{a)}.}
\end{table*}
The phenomenon of single-photon superradiance has recently been experimentally confirmed by fluorescence quantum yield measurements \textit{in vitro} in cytoskeletal protein fibers called microtubules \cite{Babcock2024} at room-temperature thermal equilibrium. Large, helical networks of strongly fluorescent tryptophan molecules in these protein fibers promote this effect by populating the most superradiant states at the lowest Lamb shifts, thus weighting them more strongly in the thermal Gibbs ensemble. Tryptophan molecules are each well-approximated by a two-level system \footnote{The emitting, excited transition state in the tryptophan molecule is known as the $^1L_a$ transition \cite{Callis1997Trp1La1Lb,Schenkl2005,Babcock2024,Patwa2024}. There is also a $^1L_b$ excited transition state in tryptophan, but it quickly relaxes on the order of tens of femtoseconds in aqueous environments to the $^1L_a$ state, which emits on the timescale of nanoseconds for the single molecule in solution, thus sufficiently separating the timescales to consider tryptophan as a two-level system.} absorbing in the ultraviolet at a peak wavelength of about $280$ nm \cite{Callis1997Trp1La1Lb,Schenkl2005}. The decay rate from the emitting state of a single tryptophan molecule is about $0.514$ ns$^{-1}$ \cite{Schenkl2005,Callis1997Trp1La1Lb}.

The analytical solution derived in Eq.~\eqref{eq:helix_eigenvalue_real_part} can be used to provide rapid order-of-magnitude predictions for superradiance and subradiance in protein fibers whose two-level system networks can be approximated by a continuous helix. In Fig.~\ref{fig:trp_positions} we assess the suitability of such predictions compared to detailed numerical simulations for three broad classes of protein fibers: microtubules, actin filaments, and amyloid fibrils.

In Table~\ref{tab:estimates_biostructures}, we show how estimates derived from the helical solution \eqref{eq:helix_eigenvalue_real_part} compare to the values numerically calculated for helical emitter networks in microtubules, actin filaments, and amyloid fibrils. The computational models from Ref.~\cite{Patwa2024} and used here were created using crystallographic \cite{lowe2001pdb1jff} and cryogenic electron microscopy \cite{liberta2019pdb6mst, gurel2017pdb6bno} data, which resolve the positions of each atom in each protein subunit up to $3.5\text{\AA}$, $5.5\text{\AA}$, and $2.7\text{\AA}$ for the microtubule, actin filament, and amyloid fibril, respectively. From these models, it is observed that the tryptophans in each structure form helical geometries, as seen in panel \textbf{a)} of Fig.~\ref{fig:trp_positions}. We find a single-helix approximation for the tryptophan network (or a helical subset of the entire network) in each of these protein fibers, which is shown as the gray line in panels \textbf{b)}, \textbf{c)}, and \textbf{d)} from Fig.~\ref{fig:trp_positions}. For actin filaments and amyloid fibrils, in which the tryptophan networks form multiple helices, this involves considering just one of those helices to use for the estimate from the analytical solution in Eq.~\eqref{eq:helix_eigenvalue_real_part}. The parameters we use for the single-helix approximation for each structure can be found in Table~\ref{tab:estimates_biostructures}. 

The maximal decay rate occurs when $\kappa=1$ in Eq.~\eqref{eq:helix_eigenvalue_real_part}, as discussed in Section \ref{sect:decay_rate_enhancements}. Since all the $\Omega$ values in Table~\ref{tab:estimates_biostructures} are greater than $2$, the interval $I$ from Eq.~\eqref{eq:interval_I} will have a range less than $1$, so only one term is present in the decay rate sum of Eq.~\eqref{eq:helix_eigenvalue_real_part}. The maximum term is the $m=0$ term, and if $\kappa=1$ the Bessel function will be $J_0(0)^2=1$. So, $\Gamma_{\text{max}}/\gamma$ is simply equal to the prefactor $2\pi n_0/k_0$, where $n_0$ is the line density. This is how the ``$\Gamma_{\text{max}}/\gamma$ from helix" row in Table~\ref{tab:estimates_biostructures} is calculated. We can also see from this result that the maximally superradiant state scales linearly with the line density $n_0$ with a constant of proportionality $2\pi/k_0$.

The eigenspectra of structures similar to these protein fibers is displayed in panel \textbf{e)} of Fig.~\ref{fig:trp_positions}, where the $n_0$ values of each protein fiber are used to plot $\Gamma/\gamma$ and $E/\gamma$ with $\kappa$. This representation of the data gives an immediate sense of which states are superradiant and subradiant for parameter regimes similar to that of realistic protein fibers: a vertical coordinate greater (less) than $1$ indicates a superradiant (subradiant) state.

We also estimate the percent of trapped states in these protein fibers using our infinite helix solution. The percent of trapped states in the infinite helix solution is $(\Omega-2)/\Omega$ for a helix with $\Omega\ge2$ (see Section \ref{sec:trapped_states} for more details). Since exactly trapped states (with $\Gamma_j = 0$) do not exist in finite structures, we compare the fraction of \textit{trapped} states in the infinite helix solution to the fraction of \textit{subradiant} states (with $\Gamma_j/\gamma<1$) in the protein fibers from numerical simulation. The agreement is within $\sim6\%$ and within $\sim21\%$ \footnote{We calculate percent difference between $v_1$ and $v_2$ as $\%_{\text{diff}}=|v_1-v_2|/\bar{v}$, where $\bar{v}=(1/2)(v_1+v_2)$.} for a microtubule and for an actin filament (single-helix), respectively, but it is much worse for an amyloid fibril, for two reasons. The first is that, as $\Omega \rightarrow 0$, the infinite continuous helix decay rates approach those of the infinite continuous line, which have trapped states for all $|\kappa|>1$. However, the percentage of measure-one regions of trapped states for the infinite continuous helix when $\Omega \rightarrow 2$ approaches $0$, and is exactly 0 for $0 < \Omega \leq 2$. So the formula $(\Omega-2)/\Omega$ is not an accurate reflection of realistic helical proteins that exhibit relatively long distances between helical turns, such as in this type of amyloid fibril. The second reason is that our infinite continuous helix model only incorporates long-range interaction terms that scale as the inverse of the distance between quantum emitters ($1/r$). But in this amyloid fibril, consecutive emitters are very close to one another ($0.5$ nm) compared to their peak excitation wavelength ($280$ nm), so the short-range terms of order $1/r^2$ and $1/r^3$ play an outsized role. Neglecting them in our infinite continuous helix model for such densely spaced protein networks of these quantum emitters thus represents an extremely rough approximation.

For the microtubule and the amyloid fibril, we find agreement within a factor of about $1.5-3$ between $\Gamma_{\text{max}}/\gamma$ from the helix and $\Gamma_{\text{max}}/\gamma$ from the protein, which is excellent given the numerous differences between a biological protein and an idealized, infinitely long helix with a continuous distribution of quantum emitters. This time, the prediction is worse for the actin filament. This is because the orientations of each quantum emitter in the actin filament, specified by the transition dipole vector, are arranged such that the superradiance is severely dampened, more so than for the microtubule and amyloid fibril. The analytical helix solution \eqref{eq:helix_eigenvalue_real_part} does not take the transition dipole vectors into account, as stated in Section \ref{sect:discrete_line_vs_continuous_line}, but the numerical approach in Ref.~\cite{Patwa2024} does. In Ref.~\cite{Patwa2024}, the interaction between the transition dipole vector network and electromagnetic field is modeled with a non-Hermitian effective Hamiltonian, which is derived in the Lindblad equation framework from Ref.~\cite{AAG2017} (described at the end of Section \ref{sect:discrete_line_vs_continuous_line}). 

Accurate estimation of superradiant and subradiant states in realistic protein fibers is advancing our understanding of the potential role of quantum-enhanced photoprotection and information processing in both neural \cite{Patwa2024} and aneural \cite{Bajpai2025} living systems.

\section{Discussion} 
That changing the geometry of the helix (specifically, the pitch) modulates how trapped eigenstates exist as a set of measure one is an interesting feature of a helical system, not present in previously studied geometries \cite{Svidzinsky2016} such as a cylinder or spheroid, where there are only trapped states for $\kappa>1$ regardless of the geometric parameters. Thus, helical architectures of quantum emitters can be used to generate tailored subradiant quantum memories \cite{AAG2017} that store information for long times and couple distinctly to electromagnetic field readouts, controlling \textit{at which} collective wavenumber $\kappa$ the subradiant states occur. Similar protocols have been developed for subradiant engineering, such as using an electric field gradient to imprint linearly increasing phases on $N$ two-level systems interacting with multiple photons \cite{Jen2017PhaseImprinting}. Our findings for helical systems present ways to achieve subradiant quantum memories in the single-photon excitation manifold by purely modifying the geometry rather than the field gradient.

When calculating the eigensolutions for the infinite continuous cylinder, helix, and line, we solve for the eigenfunctions $\beta(t, \rvec)$ in Eq.~\ref{eq:form-of-beta}. These eigenfunctions serve as coefficients in the continuous \textit{site} basis. In other words, the norm of $\beta(t, \rvec)$ is the probability amplitude of finding an excitation at the position $\rvec$ and at the time $t$. Scully \cite{Scully2009} used the so-called timed-Dicke basis, which contains phase factors $\exp(i\vec{k}\cdot \rvec_j)$ on various terms (see Table 1 in \cite{Scully2009}), to solve for the collective Lamb shifts and decay rates for a large cloud of $N$ quantum emitters in the single-photon limit. He showed that for a large sample much bigger than the excitation wavelength, the timed-Dicke state decays rapidly to the ground state, while suggesting that the \textit{symmetric} Dicke state transfers excitation more rapidly to the subradiant states, rather than the ground state. In other words, for sample sizes larger than the excitation wavelength, the timed-Dicke state is superradiant, while the symmetric Dicke state is effectively subradiant \cite{SvidzinskyScully2009}.

The decay rate for the timed-Dicke state found in Ref.~\cite{Scully2009} for a \textit{spherical} cloud of quantum emitters of radius $R$ is $\frac{3}{2(k_0R)^2}N\gamma$, whereas the timed-Dicke state $\exp(i k_0 z)$ in our infinite continuous helix decays at a rate $\frac{2\pi}{k_0}n_0\gamma$, with $n_0$ the line density of emitters. So, the characteristic maximal decay rate scaling is present in both expressions, but the helical geometry changes the result in unique ways. For example, the geometry of the helix is encoded in $n_0$, which is defined in the continuous case as $dN/dl$: the infinitesimal number of quantum emitters $dN$ in an infinitesimal length $dl$. For a helix, $dl=\sqrt{R^2 + dz^2}$. A specific helix parameterization will introduce the pitch $b$ in the infinitesimal $dz$. Thus, even though the maximal decay rate for an infinite helix scales as $(k_0R)^{-1}$, it has an additional geometric parameter (namely, $z$) that alters the coefficient of this scaling compared to the spherically symmetric case studied in Ref.~\cite{Scully2009}, where the maximal decay rate depends on $(k_0R)^{-2}$. 

Interplay between the timed-Dicke and symmetric Dicke states has been demonstrated in experimental schemes showing ultrafast switching between the two \cite{Scully2015}, as well as in an experimental demonstration of short superradiant emission and subsequent long-time oscillation in a pure Bose-Einstein condensate (BEC) of $\sim10^5$ rubidium atoms \cite{Mi2021}. In the BEC experiment, the exponential decay from the superradiance lattice is temporally separable from the long-time oscillating decay due to population transport from subradiant states, manifesting the energy-band structure and providing signatures of subradiant timed-Dicke states. Our result showing that changes in the helical pitch affect the collective decay rates of individual states might be used in a similar protocol; for example, a state with a given $\kappa$ could be rapidly switched from a trapped to a superradiant state by flexibly and reversibly changing the helical pitch.

The effect of chirality on the collective decay rate has been studied in Refs. \cite{Patwa2024} and \cite{Yelin2024}. Ref. \cite{Patwa2024} studies cylindrical arrangements of vectorial two-level systems. It is shown in Ref. \cite{Patwa2024} that if the transition dipole vectors are oriented in an achiral arrangement (for example, all pointing along the longitudinal axis of the helix), they exhibit reduced thermal quantum yields (from thermally averaged decay rates) with increasing lengths, while chiral arrangements of the transition dipole vectors lead to greater thermal quantum yields with increasing lengths. Ref. \cite{Yelin2024} considers a collection of \textit{three}-level systems, each with two excited-state levels in a $V$ configuration. Each of the two excited states is excited by only either right- or left-circularly polarized light. In this system, the authors found that helix chirality and light polarization affect which states are superradiant. This can be seen in their Fig. 4b) in Ref. \cite{Yelin2024}, where brighter portions near the ends of a finite helix indicate enhanced superradiant contributions to the emitted electric field intensity from the two photonic spin states.

These two effects (enhanced thermal quantum yield by a chiral arrangement of quantum emitters and preferential radiation of photons due to helix chirality and polarization) occur due to a chiral structure in the quantum emitter architecture. In Ref. \cite{Patwa2024}, only the collective architecture has chirality, while in the case of Ref. \cite{Yelin2024}, both the quantum emitter and the architecture have chirality. The results in Ref. \cite{Yelin2024} depend on the fact that their two excited states are selectively excited by right- or left-circularly polarized light, respectively, which is not considered in Ref. \cite{Patwa2024}. In this work, our infinite continuous helix of scalar two-level systems exhibits collective decay rate enhancements in the thermal ensemble, similar in magnitude to what we have calculated for the infinite continuous cylindrical surface (data not shown). This is unsurprising considering that for some values of $\kappa$, their solutions match in the appropriate limits when the helical pitch becomes smaller than the excitation wavelength (see Section \ref{sect:cylinder_helix_comparison} for more details). However, as Ref.~\cite{Patwa2024} makes clear, a full treatment of vectorial emitters for the continuous cases will likely exhibit sensitive dependencies on the chiral architecture and the light polarization. 

An important feature of the infinite helix eigenvalues, which is also present in the infinite cylinder and infinite lines, is the emergence of maximal decay rate states with collective Lamb shifts diverging to $-\infty$. Large-magnitude, negative collective Lamb shifts have been shown experimentally to enhance the robustness of the fluorescence quantum yield (which measures the fraction of absorbed photons emitted radiatively to the field) in room-temperature protein fibers at thermal equilibrium \cite{Babcock2024}. Thus, accurate estimations of the brightness of these superradiant states with analytical solutions such as the infinite helix presented here can help guide experiments for other superradiant systems in flexible biomaterials. There is also vast potential for superradiant states to be used in information processing, since these states in microtubules have lifetimes that support logical operations between orthogonal states over a billion times faster than the computational speed of a single Hodgkin-Huxley neuron, and within just a few orders of magnitude of the Margolus-Levitin limit for UV excitations \cite{Kurian2025}. 

Another intriguing application for these collective light-matter interactions is the direct detection of single photons by the human eye. This detection involves a priming process, suggested by the fact that the probability of subjects correctly reporting a single photon is enhanced by the arrival of an earlier photon within about a $5$-second time interval \cite{tinsley2016direct}. Such an observation raises questions about the mechanisms of single-photon signal propagation, post-processing, and thresholding in rod photoreceptors, which have been reported to detect a few photons (typically less than 10) during an integration time of 300 ms \cite{hecht1942energy, barlow1956retinal, okawa2007optimization}. As each single photon incident on the eye does not produce a retinal isomerization event and subsequent signal transduction cascade, this disparity in timescales between the reported integration time of the human visual system \cite{field2005retinal} and the long priming process should motivate more detailed analyses of the subradiant properties of helical architectures of the retinal chromophore (in the protein rhodopsin) in rod photoreceptors. Using existing setups to generate quantum light and preparations of single photons in various superposition states, theoretical proposals for entanglement detection with the naked eyes of human observers \cite{brunner2008possible}---assuming photoresponse functions smoother than ideal steps and under postselection for conclusive events at both observers---may be closer to realization than ever before.

\section{\label{sect:conclusion}Conclusions}
In this work, we studied the interaction of a continuous, infinite helical distribution of two-level emitters with a single photon, found a novel analytical eigensolution, and compared it to the helix's topological (one-dimensional) equivalent: a continuous, infinite straight line of emitters. To do this, we utilized the formalism of Ref.~\cite{Svidzinsky2016} which describes the light-matter interaction of continuous emitter distributions on surfaces/shells with a single photon. We make a detailed comparison of our solutions and two other analytical solutions: an infinitely long cylindrical surface, and an infinite, discrete line of quantum emitters. For the infinite helix, we described in detail the complex relationship between its geometric parameters and the presence of superradiant and trapped states. The tuning of helix parameters---the pitch $b$ and radius $r$---can be used to control where superradiant and subradiant states occur in the eigenspectrum. The insight from this analysis was used to make order-of-magnitude predictions of superradiance and subradiance of quantum emitter architectures present in realistic protein fibers. We find that when the inverse pitch $\Omega=2\pi/k_0 b$ and/or the dimensionless radius $r=k_0 R$ are very small, the thermally averaged decay rate is enhanced. The parameter regimes of the helical quantum emitter arrangements in the protein fibers do have very small values of $r$, which suggests that the structure of these protein fibers are finely tuned to maximize quantum-optical effects in the thermal average. Experimental observation of single-photon superradiance in microtubules also supports this fact \cite{Babcock2024}, and experiments are ongoing to confirm this result in helical emitter architectures in other protein fibers \cite{Patwa2024}.

We also investigated the interplay between the number of geometric free parameters in these structures. The infinite continuous line has 0, the infinite cylindrical surface has 1 (the radius $r$), and the infinite continuous helix has 2 (the helical pitch $b$ and the radius $r$). The infinite discrete line, on the other hand, has two extra geometric degrees of freedom per emitter, since each emitter has an orientation specified by its transition dipole unit vector and for which normalization exactly specifies the other unit vector component. If we assume that all the quantum emitters are oriented in the same direction, as is assumed in the analysis of the discrete infinite line in Ref.~\cite{AAG2017}, then only two more free parameters (instead of $2N$) are added. This makes the total number of geometric free parameters equal to three for the discrete line (the spacing $d$ between emitters and two components of the transition dipole unit vector) and five for the discrete helix ($b$, $r$, $d$, and the two free transition dipole components).

The topological dimension of each structure and the number of variables needed to parameterize it are of course distinct from the dimension of the space in which it is embedded. The infinite continuous line and helix are both topologically equivalent, of dimension 1, to the reals ($\mathbb{R}$), while the infinite discrete line and helix are topologically equivalent, of dimension 0, to the integers ($\mathbb{Z}$). The infinite continuous cylinder is topologically equivalent to $\mathbb{R} \times S^2$, where $S^2$ reflects a 2-sphere (circle) in the higher embedding of 3-dimensional space. Only the helix shares the property of the cylinder, also being embedded in 3-dimensional space, rather than the 1-dimensional embeddings for the lines. In principle, one can create a non-trivial ``topological edge" (analogous to a cylinder's) from a helix embedded in 3D, which one cannot create from a line.

We found that topologically equivalent structures can exhibit different quantum optical properties. For example, the line and the helix are topologically equivalent, but the fact that the infinite continuous helix is embedded in 3D creates additional features not present in the infinite continuous line, such as non-trapped states at $k_z>k_0$ (or, equivalently, $\kappa>1$). A consequence of the 3D embedding is the fact that the infinite or finite helix has a nonzero winding number about the center of the helix when looking down its longitudinal axis and projected onto the plane, while the infinite line has $0$ winding number. Such a nonzero winding number is only possible for a topologically one-dimensional object if it is embedded in more than one spatial dimension. Moreover, while the helix and cylinder are topologically distinct, there are certain limits in which the decay rates of these two are equivalent, even though the geometric parameters required for the helix are double that of the cylinder. 

The differences we find between our infinite continuous helix and line solutions are in contrast to topological invariants found in other non-Hermitian open quantum systems \cite{Yao2018Topological,Song2019Topological}. Specifically, we do not see something analogous to the non-Hermitian skin effect, in which there are only discretely many spatially delocalized eigenstates, and the rest are bunched up near or along the boundaries of the structure. This skin effect occurs due to open boundary conditions and finite structures. Finite-length effects and broken symmetries other than the skin effect, which can manifest as topologically non-trivial eigenstate delocalization patterns, have been described with detailed numerical simulations of single-photon superradiant states and their probability amplitudes across the emitter site basis in finite vibrating microtubules with slight deformations (see the supplementary information of Ref.~\cite{Patwa2024}). In our infinite continuous helix and line solutions, there are no boundary conditions at the topological edges because of the infinite length of the structures, so fully delocalized eigenstates of the form $\exp(ik_z z)$ emerge and the non-Hermitian skin effect is not observed. Furthermore, the edge associated with the radius $R$ in discrete or continuous spheres, cylinders, and helices is clearly periodic, while open boundary conditions are crucial to the skin effect, requiring full-fledged Lindblad dynamics beyond the effective non-Hermitian Hamiltonian for an open quantum system.

Topological arguments are also revealing when applied to realistic, discrete protein fiber architectures, such as the microtubule. An alternative to considering the tryptophan network in a microtubule as a helix, shown in Fig.~\ref{fig:trp_positions}b, is to approximate the structure as a cylindrical lattice of emitters. Such a lattice is isomorphic to a 2D discretized grid/plane $\mathbb{Z}_m\times\mathbb{Z}_n$, with periodic boundary conditions in the ``wrapped'' dimension, which is isomorphic to the group $(\mathbb{Z}_{mn},+)$, the integers from $0$ to $mn$ with addition modulo $mn$. Of course, the rectangular grid with $m \times n$ emitter sites that maps to the finite cylindrical lattice has a width equal to the cylinder base circumference, and height equal to the cylinder length. Interestingly, it has been shown that in microtubules only when this cylinder length approaches or exceeds the cylinder base circumference (i.e., only when the height of the rectangle is approximately equal to or greater than its width) do superradiant states at the lowest-lying energy, with largest negative collective Lamb shifts, emerge \cite{Celardo2019}. This feature of microtubules guarantees that helical-cylindrical microtubule architectures will exhibit extraordinarily robust signatures of single-photon superradiance in their steady-state thermal quantum yields of fluorescence \cite{Babcock2024}.

By comparing structures with varying geometric free parameters, topological dimension, and embedding space dimension, we hope to decrease the gap between formalisms that operate with structures of different dimension (of topology and/or embedding space) and measure (continuous vs. discrete). Future work may include extending the approach of Ref.~\cite{AAG2017} to continuous structures, and comparing it to our results for the infinite continuous line and helix in this paper. It would also be interesting to extend the approach of Ref.~\cite{Svidzinsky2016} to include polarization effects for a better comparison with our solution for the infinite discrete line. These studies may lead to an overarching theoretical bridge between diverse approaches to the collective emission and storage of light in quantum matter.

\section*{Acknowledgments}
This research was supported by the Alfred P. Sloan Foundation, the Chaikin-Wile Foundation, and the National Science Foundation (NSF). The authors would also like to acknowledge insightful conversations with Anatoly Svidzinsky and Marlan Scully that took place at the Princeton-Texas A\&M University Quantum Summer School in Wyoming; with Ana Asenjo-Garcia and Darrick Chang at the Kavli Institute for Theoretical Physics (KITP) in California; and with Luca Celardo over the years. KITP is supported by a grant from the NSF. High performance computing resources from the Oak Ridge and Argonne Leadership Computing
Facilities made the numerical simulations of superradiance in large protein systems \cite{Babcock2024,Patwa2024} possible.

\bibliographystyle{unsrt}
\bibliography{citations}

\end{document}